\documentclass[aps,physrev,amsmath,amssymb,reprint]{revtex4-2}

\usepackage{graphicx}
\usepackage{xcolor}
\usepackage{floatrow}
\usepackage[caption=false,subrefformat=parens,labelformat=parens]{subfig}
\usepackage{dcolumn}

\usepackage{bm}
\usepackage{mathrsfs}
\DeclareMathOperator{\csch}{csch}
\usepackage{siunitx}

\usepackage[colorlinks=true,linkcolor=blue,citecolor=blue,urlcolor=blue]{hyperref}

\begin{document}

\title{Cross-linked pair of polymer chains under strong tension}

\author{Geunho Noh}
\email[]{ngh@knu.ac.kr}
\author{Panayotis Benetatos}
\email[]{pben@knu.ac.kr}
\affiliation{Department of Physics, Kyungpook National University, Daegu 41566, Republic of Korea}

\date{\today}

\begin{abstract}

We study two cross-linked polymer systems in the strong stretching regime. The first consists of two polymers sharing one endpoint, with the other two endpoints coupled by a harmonic potential. Within the weakly bending approximation, we analyze the tensile elastic response for freely jointed or wormlike chains; for the latter, the approximation applies either at large tension or at moderate tension with large persistence length (rodlike limit). We obtain analytic expressions for the force--extension relation and for the longitudinal and transverse mismatch of the cross-linked endpoints. In the thermodynamic limit, the cross-link does not affect the tensile elasticity, but it significantly suppresses transverse fluctuations, effectively forming a loop structure. The second system is a polymer necklace in the thermodynamic limit, composed of two strongly stretched polymers interconnected by a regular sequence of reversible cross-links. Using an analogy with a two-dimensional system of concatenated Gaussian loops (``Gaussian slinky''), we calculate the mean fraction of cross-linked sites as a function of the tensile force and find weak and strong binding regimes connected by a crossover. For shallow binding potential wells (compared with $k_{\rm{B}}T$), we employ a continuum description and exploit the mapping between directed polymers and a two-dimensional quantum particle to determine the crossover behavior and the mean transverse separation between the two polymer chains.

\end{abstract}

\maketitle

\section{Introduction}\label{sec1}

Cross-linked structures of macromolecular chains are ubiquitous in polymer networks---whether natural or synthetic, and either chemical (covalent, permanent) or physical (non-covalent, reversible)---including systems such as elastomers, plastics, and gels \cite{Johnson,Patrickios,Seiffert}. In biological systems, similar network architectures are commonly found in the cytoskeleton and extracellular matrix, where filamentous biopolymers are coupled via specific linker proteins that mediate binding interactions \cite{Burla,Fletcher,Mouw}. In particular, these biopolymers often organize into aligned bundles or parallel arrays, serving diverse structural and mechanochemical functions. For example, actin filaments form cross-linked bundles both in filopodia---protrusions that explore the cell's environment and initiate focal adhesions enabling mechanical communication and driving cell migration---and in contractile stress fibers connected to those adhesions \cite{Mattila,Blake,Burridge,Naumanen}. Collagen fibrils, in turn, assemble into hierarchical bundles that form tendons transmitting mechanical forces from muscles to bones \cite{Birk,Kong,Eekhoff,Ellingson}. Across their diversity, cross-linked and aligned biopolymer systems consistently experience mechanical stresses arising from both external loading and internal force generation \cite{Luciano,Ingber}. Their collective response is governed by the cross-linked architecture, which controls mechanical properties such as the elastic modulus \cite{Gardel,Piechocka} and bending stiffness \cite{Claessens,Heussinger1,Bathe,Heussinger2}.

Although the collective mechanical behavior of these filamentous assemblies arises from complex interfilament coupling and environmental variability, it is instructive to examine minimal, isolated model systems that capture the essential physics and provide insight into the origin of the overall behavior. At the individual-filament level, the constituent chains are typically semiflexible, behaving as smooth, homogeneous elastic rods that resist bending and exhibit conformational fluctuations well described by the wormlike chain model \cite{Kratky-Porod,Saitô,Frey}. From a bottom-up perspective, the most elementary nontrivial building block of such assemblies is a pair of cross-linked semiflexible (wormlike) chains. Unidirectional alignment can be achieved by imposing sufficiently strong tension such that the chains become effectively oriented along the fixed tensile-force axis. Under this tension-induced entropic confinement, their preferred orientation allows the filaments to be treated as directed polymers in the weakly bending (tilting) regime, whose configurations are dominated by small transverse fluctuations about the force axis \cite{deGennes1,deGennes2,Huse-Henley,Kaimien,Nelson1,Marko-Siggia}. The weakly bending regime may also be realized in stiff filaments with large persistence length---an intrinsic measure of polymer stiffness defined as the exponential decay length of orientational correlations along the contour---where the filaments remain nearly straight even without applied tension. However, in order for a pair of chains to maintain a mutually aligned configuration with a well-defined common axis, a sufficiently strong tensile force is still required to suppress large orientational deviations (relative tilting) between the two chains.

Cross-linked polymer pairs under tension and related directed-chain models have been studied theoretically, elucidating their conformational statistics, elastic response, and binding behavior. In Ref.~\cite{PB1}, the tensile elasticity of randomly cross-linked polymer bundles was formulated on the basis of the force-extension relation at the pair level. Under strong stretching, the cross-links contribute to tensile stiffening, independent of the bending stiffness of the chains. A similar effect of cross-links on the tensile elastic behavior was reported in Ref.~\cite{Alice}, where two weakly bent wormlike chains are connected by an arbitrary number of cross-links. In that model, the cross-links enhance the tensile stiffness of the chain pair, which increases with both the number and the strength of the cross-links. Whereas the preceding studies considered permanently cross-linked systems, with the cross-links represented by harmonic springs, Ref.~\cite{PB2} addressed reversible harmonic cross-links and their tension-induced binding behavior, extending the irreversible model of Ref.~\cite{Alice}. Within a mean-field framework, in which fluctuations in the number of cross-links are neglected and replaced by their average value, two distinct binding regimes emerge, exhibiting a force-induced crossover. The same study also analyzed a related model of a single directed polymer in a continuous potential well, drawing an analogy to the crossover between weak and strong localization of a vortex line in a type-$\mathrm{II}$ superconductor subject to a columnar pinning potential \cite{Nelson1,Nelson2}.

The binding behavior of a cross-linked polymer pair is closely related to the melting (denaturation) of double-stranded DNA, in which the two strands are locally and transiently held together by reversible base-pair interactions mediated by hydrogen bonds. One of the simplest theoretical descriptions of DNA melting, encompassing both thermal and mechanical denaturation, is the classical Poland--Scheraga model~\cite{Poland-Scheraga1} and its various extensions~\cite{Fisher1,Kafri1,Kafri2,Metzler1,Metzler2,Rudnick1,Rahi,NGH1,PB3,NGH2}, which rest upon the generating function framework for alternating sequences of two segment types~\cite{Lifson,Litan,Roe1,Roe2,Poland-Scheraga2}, such as bound (intact base pairs) and unbound (denatured loops) regions in DNA. For those variants that incorporate external tension, the chains have been modeled in distinct ways---ranging from rigid rods and random walks with self-avoiding interactions~\cite{Metzler1,Metzler2}, to self-avoiding freely jointed chains with tunable stiffness~\cite{Rudnick1}, semiflexible chains with bending-energy penalties~\cite{Rahi}, freely jointed chains with two-state hinges~\cite{NGH1}, wormlike chains with two-state bending stiffness~\cite{PB3}, and flexible chains and loops with different degrees of stiffness~\cite{NGH2}. Such distinct representations influence the qualitative binding behavior, affecting the existence and the nature of possible transitions, as well as the topology of the resulting phase diagram.

In this work, we develop a theoretical framework for a cross-linked polymer pair under strong tension, focusing on two main aspects: the effect of a single cross-link connecting the end-points of the two chains on the tensile elastic response, and the force-dependent binding behavior of a sequence of reversible cross-links. While our primary motivation stems from semiflexible biopolymer systems, we also analyze the same aspects within the freely jointed chain model on equal footing. This model is included not only for its theoretical interest but also for its relevance, as it serves as a precursor to the wormlike-chain counterpart and provides conceptual continuity.

This article is organized as follows. Section~\ref{sec2} begins with a cross-linked freely jointed chain pair, which we refer to as a loop to emphasize its looped configuration rather than a true topological loop. In the strong stretching regime, we construct the partition function and evaluate the force--extension relation as well as the conformational fluctuations. From the analysis of the transverse fluctuations, we justify our notion of the loop formed by the chain pair. Guided by the formulation of the cross-linked freely jointed loop, we examine in parallel, in Sec.~\ref{sec3}, the cross-linked wormlike loop in the weakly bending regime. Unlike the freely jointed case, the tensile elastic behavior of the wormlike model requires a more intricate analysis, as three characteristic length scales---the contour length, persistence length, and memory length (defined in that section)---enter the description and delineate distinct regimes of elastic behavior. In Sec.~\ref{sec4}, we turn to the second main topic and consider a necklace-like polymer system---a long polymer pair interconnected by multiple reversible cross-links. Within both the freely jointed and wormlike chain models, we revisit the statistics of transverse fluctuations of a single chain in the strong stretching regime and evaluate the mean occupation number of bound cross-links in the thermodynamic limit. We also formulate a complementary approach to analyze the binding behavior of the necklace in a continuum limit of the cross-linking interaction, employing a quantum mechanical analogy inherent to directed polymer systems. Finally, in Sec.~\ref{sec5}, we summarize the main results of this work.

\section{Cross-linked Freely Jointed Loop}\label{sec2}

\subsection{Model}

\begin{figure}
    \includegraphics[width=8.6cm]{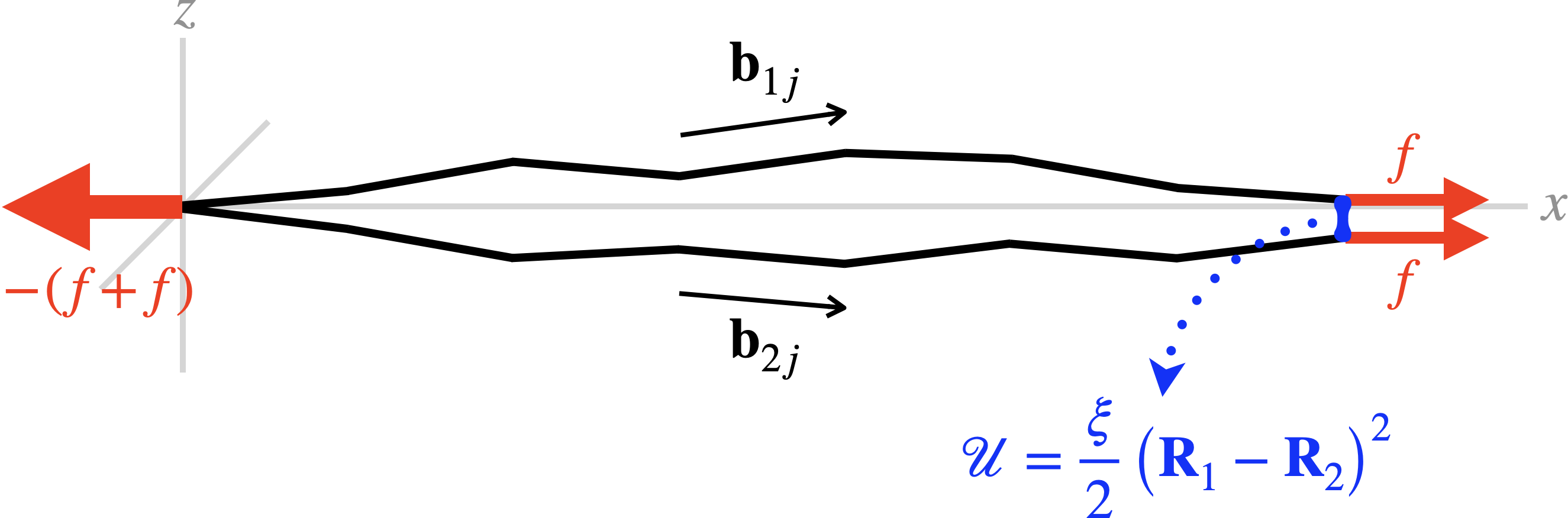}
    \caption{Schematic diagram of the cross-linked freely jointed loop under strong tension.}
    \label{fig:cFJL}
\end{figure}

We first consider two freely jointed chains, each consisting of $N$ rigid links of equal length $b$, as illustrated in Fig.~\ref{fig:cFJL}. The sequence of $N$ segments is represented by the set of bond vectors $\{\bm{b}_{ij}\}$, where $i$ is the chain index ($i=1,2$) and $j$ is the bond (link) index ($j=1,\cdots,N$). The end-to-end vector of chain $i$ is expressed as $\bm{R}_i=\sum_{j=1}^N\bm{b}_{ij}$. One endpoint of each chain is fixed at the origin of a three-dimensional reference frame, while the other endpoints are coupled by a single cross-link. We model the cross-link as a harmonic potential, $\mathscr{U}=\xi (\bm{R}_1-\bm{R}_2)^2/2$, where $\xi$ is the spring constant. We refer to this system as the cross-linked freely jointed loop (cFJL). When each chain of the cFJL is under tension by a pair of stretching forces $\pm f\hat{x}$ applied at its ends, the system is described by the effective Hamiltonian that comprises the two types of potential energy:
\begin{equation}\label{H_cFJL}
    \mathscr{H}_{\rm{cFJL}}=\frac{\xi}{2}\left(\sum_{j=1}^{N}(\bm{b}_{1j}-\bm{b}_{2j})\right)^2-f\sum_{j=1}^{N}(b_{1j,x}+b_{2j,x}),
\end{equation} 
where $b_{ij,x}$ denotes the $x$-component of $\bm{b}_{ij}$.

In the strong stretching regime, where the bond vectors are nearly aligned with the force axis, the local inextensibility constraint ($\lvert\bm{b}_{ij}\rvert=b$) provides a second-order approximation for the longitudinal component in terms of the transverse components:
\begin{equation}\label{b_ij,x}
    b_{ij,x}\approx b\left(1-\frac{{\bm{b}_{ij,\perp}}^2}{2b^2}\right),\quad\lvert\bm{b}_{ij,\perp}\rvert \ll b,
\end{equation}
where $\bm{b}_{ij,\perp}\equiv(b_{ij,y},b_{ij,z})$ denotes the two-component transverse bond vector. Substituting Eq.~\eqref{b_ij,x} into Eq.~\eqref{H_cFJL} and keeping terms up to quadratic order, we obtain the effective Hamiltonian in the strong stretching regime:
\begin{align}\label{H_cFJL'}
    \mathscr{H}_{\rm{cFJL}}&=\frac{\xi}{2}\left(\sum_{j=1}^{N}(\bm{b}_{1j,\perp}-\bm{b}_{2j,\perp})\right)^2-Nb(f_1+f_2)\nonumber\\
    &\quad+\frac{1}{2b}\sum_{j=1}^N\left(f_1{\bm{b}_{1j,\perp}}^2+f_2{\bm{b}_{2j,\perp}}^2\right).
\end{align}
Here, we deliberately label the applied forces on the two chains as $f_i$ to extract the desired observables later via force differentiation. This labeling is also useful for treating the general case of asymmetric stretching ($f_1\neq f_2$). In this paper, however, we focus our analysis on the symmetric case ($f_1=f_2$). It should be noted that the Hamiltonian in Eq.~\eqref{H_cFJL'} does not capture the cross-linking interaction arising from the longitudinal mismatch of the endpoints, as the corresponding leading-order term is quartic. Consequently, this formulation is essentially limited to transverse coupling, under the assumption that longitudinal contributions are negligible in the strong stretching regime.

\subsection{Partition function}

We now construct the canonical partition function of the strongly stretched cFJL. Because the two transverse directions are independent and contribute identically, it suffices to analyze a single-component partial Hamiltonian
\begin{align}\label{H_cFJL*}
    \mathscr{H}_{\rm{cFJL}}^{*}&\equiv\frac{\xi b^2}{2}\left(\sum_{j=1}^{N}(u_{1j,\perp}-u_{2j,\perp})\right)^2\nonumber\\
    &\quad+\frac{b}{2}\sum_{j=1}^N\left(f_1{u_{1j,\perp}}^2+f_2{u_{2j,\perp}}^2\right),
\end{align}
where the variable $b_{ij,\perp}$ has been rescaled into dimensionless form $u_{ij,\perp}\equiv b_{ij,\perp}/b$. In matrix notation, this expression can be written more succinctly as
\begin{equation}\label{H_cFJL*_matrix}
    \frac{\mathscr{H}_{\rm{cFJL}}^{*}}{k_{\rm{B}}T}=\frac{1}{2}{\mathbf{u}}^{\mathsf{T}}\left(\mathbf{v}\mathbf{v}^{\mathsf{T}}+\mathbf{\Lambda}\right)\mathbf{u},
\end{equation}
where
\begin{equation*}
    \mathbf{u}\equiv\begin{bmatrix}
                        u_{11,\perp}\\
                        u_{21,\perp}\\
                        \vdots\\
                        u_{1N,\perp}\\
                        u_{2N,\perp}
                    \end{bmatrix}_{2N\times1},
    \quad  
    \mathbf{v}\equiv\sqrt{\frac{\xi b^2}{k_{\rm{B}}T}}
                \begin{bmatrix}
                    1\\
                    -1\\
                    \vdots\\
                    1\\
                    -1
                \end{bmatrix}_{2N\times1},
\end{equation*}
\begin{equation*}
    \mathbf{\Lambda}\equiv\frac{b}{k_{\rm{B}}T}
                \begin{bmatrix}
                    f_1 & 0 & \cdots & 0 & 0\\
                    0 & f_2 & \cdots & 0 & 0\\
                    \vdots & \vdots & \ddots & \vdots & \vdots\\
                    0 & 0 & \cdots & f_1 & 0\\
                    0 & 0 & \cdots & 0 & f_2
                \end{bmatrix}_{2N\times2N}.
\end{equation*}
The corresponding partition function then follows as
\begin{align}\label{Z_cFJL*}
    \mathscr{Z}_{\rm{cFJL}}^{*}=\int d^{2N}\rm{u}\exp\left(-\frac{1}{2}{\mathbf{u}}^{\mathsf{T}}\left(\mathbf{v}\mathbf{v}^{\mathsf{T}}+\mathbf{\Lambda}\right)\mathbf{u}\right),
\end{align}
where the 2$N$-dimensional integral spans all possible values of $u_{ij}$. The integrand is sharply peaked in the 2$N$-dimensional space, reflecting the suppression of large transverse undulations in the strong stretching regime. Accordingly, Eq.~\eqref{Z_cFJL*} is well approximated by the Gaussian integral ($\int_{\mathbb{R}^{2N}}d^{2N}\rm{u}$), yielding $(2\pi)^{N}{\left[\det(\mathbf{v}\mathbf{v}^{\mathsf{T}}+\mathbf{\Lambda})\right]}^{-1/2}$ \cite{Zinn-Justin}. Applying the matrix determinant lemma \cite{Harville},
\begin{equation*}
    \det(\mathbf{v}\mathbf{v}^{\mathsf{T}}+\mathbf{\Lambda})=(1+\mathbf{v}^{\mathsf{T}}{\mathbf{\Lambda}}^{-1}\mathbf{v})\det(\mathbf{\Lambda}),
\end{equation*}
we arrive at
\begin{equation}\label{Z_cFJL*_result}
    \mathscr{Z}_{\rm{cFJL}}^{*}=\left(\frac{2\pi k_{\rm{B}}T}{b\sqrt{f_1f_2}}\right)^N\left(1+\frac{N\xi b(f_1+f_2)}{f_1f_2}\right)^{-1/2}.
\end{equation}
The total partition function is then readily assembled by recalling the full expression of the Hamiltonian:
\begin{align}\label{Z_cFJL'}
    \mathscr{Z}_{\rm{cFJL}}&=\exp\left(\frac{Nb(f_1+f_2)}{k_{\rm{B}}T}\right)\left(\mathscr{Z}_{\rm{cFJL}}^{*}\right)^2
    \nonumber\\
    &=\exp\left(\frac{Nb(f_1+f_2)}{k_{\rm{B}}T}\right)\left(\frac{2\pi k_{\rm{B}}T}{b\sqrt{f_1f_2}}\right)^{2N}\nonumber\\
    &\quad\times\left(1+\frac{N\xi b(f_1+f_2)}{f_1f_2}\right)^{-1}.
\end{align}
We use this general form only when performing force derivatives. With the tracking index removed (i.e., with tension $f$ on each chain), the partition function reduces to
\begin{equation}\label{Z_cFJL}
    \mathscr{Z}_{\rm{cFJL}}=\left(\frac{2\pi\exp(bf/k_{\rm{B}}T)}{bf/k_{\rm{B}}T}\right)^{2N}\left(1+\frac{2N\xi b}{f}\right)^{-1}.
\end{equation}
Note that when $\xi=0$, the partition function takes the form $\mathscr{Z}_{\rm{cFJL}}\vert_{\xi=0}=(\mathscr{Z}_{\rm{FJC}})^2$, where  
\begin{equation}\label{Z_FJC}
    \mathscr{Z}_{\rm{FJC}}=\left(\frac{4\pi\sinh(bf/k_{\rm{B}}T)}{bf/k_{\rm{B}}T}\right)^N\approx\left(\frac{2\pi\exp(bf/k_{\rm{B}}T)}{bf/k_{\rm{B}}T}\right)^N
\end{equation}
is the partition function of a single freely jointed chain (FJC) under tension $f$ in the strong stretching regime \cite{Rubinstein}.
  
\subsection{Average extension and fluctuations}

The average extension (end-to-end projection length along the force axis) of chain $i$ is calculated from the relation
\begin{equation}\label{X-f_general}
    \langle X_i\rangle_{\rm{cFJL}}=k_{\rm{B}}T\frac{\partial\ln{\mathscr{Z}_{\rm{cFJL}}}}{\partial f_i}.
\end{equation}
Under symmetric stretching ($f_1=f_2=f$), the two chains exhibit identical average extensions, allowing us to define the average extension of the entire system as $\langle X\rangle_{\rm{cFJL}}\equiv\langle X_i\rangle_{\rm{cFJL}}$. The result can be expressed as
\begin{equation}\label{X-f_cFJL_full}
    \langle X\rangle_{\rm{cFJL}}=\langle X\rangle_{\rm{FJC}}+\langle\Xi\rangle_{\rm{cFJL}},
\end{equation}
where
\begin{equation}\label{X-f_FJC}
    \langle X\rangle_{\rm{FJC}}=Nb\left(1-\frac{k_{\rm{B}}T}{bf}\right)
\end{equation}
is the average extension of a single freely jointed chain in the strong stretching regime, and
\begin{equation}\label{correction_cFJL}
    \langle\Xi\rangle_{\rm{cFJL}}=\frac{Nk_{\rm{B}}T\xi b}{f^2+2N\xi bf}
\end{equation}
is the additional contribution due to the cross-linking interaction in the cFJL model. When $\xi=0$, the correction term vanishes, as expected and required.

From a practical perspective, the cross-link in this model does not significantly affect the tensile elasticity of the strongly stretched cFJL, as the correction term scales as $\sim1/f^2$. Even in the large $\xi$ limit, it contributes to the overall strain only a term $k_{\rm B}T/2Nbf$, which vanishes in the thermodynamic limit. In fact, this is the contribution obtained in the thermodynamic limit irrespective of the magnitude of $\xi$. Nevertheless, the cross-link enables the formation of a loop structure, which is clearly distinct from a double-chain structure where two single chains fluctuate independently with one end fixed. A natural follow-up question is how distinct the two configurations are. To address this, we examine the effect of the cross-linking on the longitudinal and, more pertinent to our discussion, transverse fluctuations.

We first consider the longitudinal mismatch between the upper and lower chains along the force axis, denoted by $R_{\parallel}\equiv X_1-X_2$. The magnitude of the longitudinal fluctuations is obtained as
\begin{align}\label{fluctuation_l_cFJL}
    \left\langle\left(R_{\parallel}-\langle R_{\parallel}\rangle\right)^2\right\rangle_{\rm{cFJL}}&=\left(k_{\rm{B}}T\right)^2\left(\frac{\partial}{\partial f_1}-\frac{\partial}{\partial f_2}\right)^2\ln\mathscr{Z}_{\rm{cFJL}}\nonumber\\
    &=2N\left(\frac{k_{\rm{B}}T}{f}\right)^2\left(1-\frac{2\xi b}{f+2N\xi b}\right).
\end{align}
When $\xi=0$, this expression reduces to $2N(k_{\rm{B}}T/f)^2$, which can also be derived from the partition function of a single freely jointed chain [Eq.~\eqref{Z_FJC}]. The factor of $2$ accounts for the two independent chains, each fixed at one end. The same result is also recovered as $N\to\infty$, implying that the longitudinal fluctuations become insensitive to the boundary condition (cross-linked or not) in the thermodynamic limit.

We next consider the transverse mismatch denoted by $\bm{R}_{\perp}\equiv\bm{R}_{1,\perp}-\bm{R}_{2,\perp}$, whose average vanishes, $\langle\bm{R}_{\perp}\rangle=0$, by symmetry in the $y$--$z$ plane. The magnitude of the transverse fluctuations is obtained as
\begin{equation}\label{fluctuation_t_cFJL}
    \langle{\bm{R}_{\perp}}^2\rangle_{\rm{cFJL}}=-2k_{\rm{B}}T\frac{\partial\ln\mathscr{Z}_{\rm{cFJL}}}{\partial\xi}=\frac{4Nk_{\rm{B}}Tb}{f+2N\xi b}.
\end{equation}
When $\xi=0$, the result is consistent with that obtained from the force--extension relation of a single freely jointed chain in the strong stretching regime:
\begin{equation}\label{fluctuation_t_FJC}
    \left.\langle{\bm{R}_{\perp}}^2\rangle_{\rm{cFJL}}\right\vert_{\xi=0}=2\times2b\left(Nb-\langle X\rangle_{\rm{FJC}}\right)=\frac{4Nk_{\rm{B}}Tb}{f}.
\end{equation}
Again, the factor of $2$ accounts for the contributions of the two independent chains. The presence of the cross-link suppresses the transverse fluctuations, and the suppression becomes more pronounced as the system size increases. Without the cross-link, the transverse fluctuations grow linearly with $N$, whereas with the cross-link they approach a constant value, $2k_{\rm{B}}T/\xi$. Consequently, in the thermodynamic limit, the relative magnitude of the transverse fluctuations, $\sqrt{\langle{\bm{R}_{\perp}}^2\rangle}/\langle X\rangle$, scales as $N^{-1/2}$ in the uncoupled case, whereas it scales as $N^{-1}$ in the coupled case. Therefore, the single cross-link markedly alters the nature of the transverse fluctuations and effectively pulls the two chain ends toward each other. This effect plays a key role in identifying the cFJL as a looped configuration, distinguishing it from the mere double-chain structure.

We now revisit the tensile elasticity of the cFJL from the perspective of a ``single-loop'' structure and contrast it with a single-chain counterpart. Consider the cFJL and the FJC under the same total stretching force $f$. Assuming that this force is applied symmetrically to the two arms of the loop (i.e., $f/2$ per arm), the average extension of the loop to leading order is given by
\begin{equation}\label{X-f_cFJL}
    \langle X\rangle_{\rm{cFJL}}=Nb\left(1-\frac{2k_{\rm{B}}T}{bf}\right).
\end{equation}
Compared to the single chain [Eq.~\eqref{X-f_FJC}], the loop requires twice the force to achieve the same extension. This observation can be rephrased more quantitatively in terms of the tensile compliance, defined as the slope of the curve in the $f$--$\langle X\rangle$ plane. Let $f_0$ be the applied force required to maintain a certain average extension $X_0$ for the loop. Then, the force required for the single chain to achieve the same extension is $f_0/2$. For each case, an infinitesimal change in the applied force $\delta f$ and the resulting deformation $\delta X$ from $X_0$ satisfy the following relations in the linear-response regime:
\begin{equation}\label{compliance_FJ}
    \delta X_{\rm{cFJL}}=\frac{2Nk_{\rm{B}}T}{{f_0}^2}\delta f,\quad\delta X_{\rm{FJC}}=\frac{4Nk_{\rm{B}}T}{{f_0}^2}\delta f.
\end{equation}
The compliance $\delta X/\delta f$ of the loop is half that of the single chain, indicating that the loop is effectively twice as stiff. For Gaussian chains and loops, this twofold difference in stiffness holds exactly---not only locally, but also globally \cite{NGH2}. In that case, one can draw an analogy to mechanical springs: a Gaussian loop is equivalent to two springs connected in parallel, irrespective of the system size or applied force.

\section{Cross-linked Wormlike Loop}\label{sec3}

\subsection{Model}

\begin{figure}
    \includegraphics[width=8.6cm]{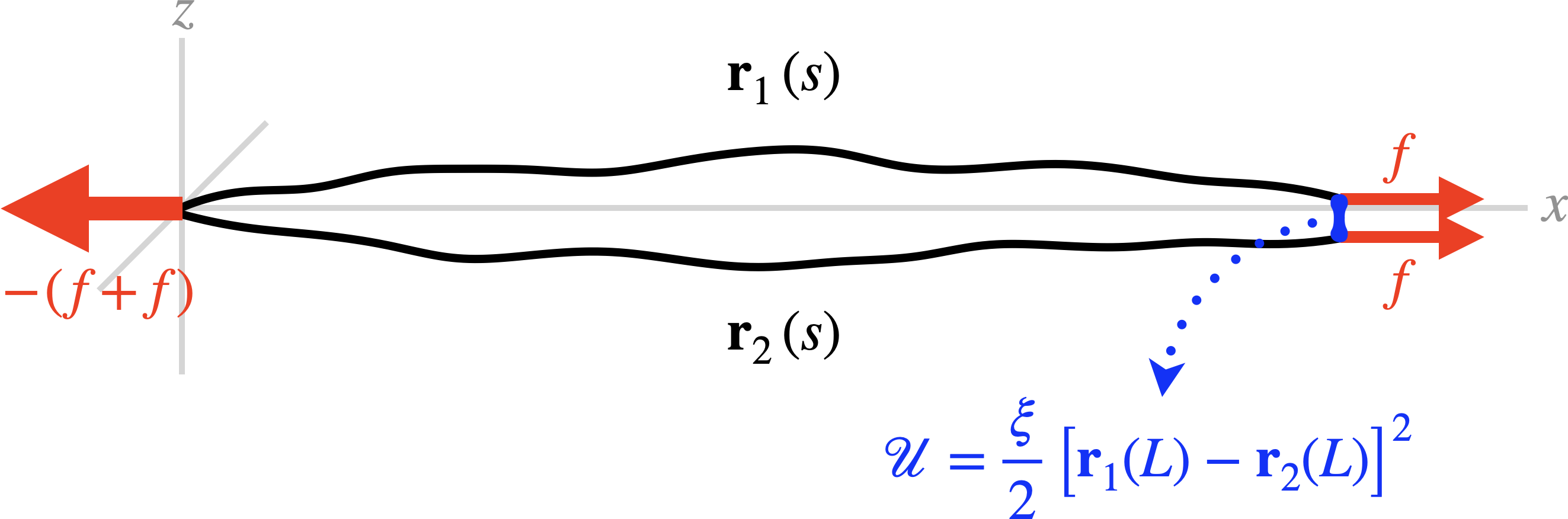}
    \caption{Schematic diagram of the cross-linked wormlike loop in the weakly bending regime, which can arise under strong tension or, for large persistence length, under moderately strong tension.}
    \label{fig:cWLL}
\end{figure}

We now turn to the cross-linked wormlike loop (cWLL), the semiflexible counterpart of the cross-linked freely jointed loop (cFJL), as illustrated in Fig.~\ref{fig:cWLL}. The cWLL consists of two wormlike chains with contour length $L$ and persistence length $l_{\rm{p}}$. Each chain is represented by a parametric curve $\bm{r}_i(s)$ in three-dimensional space, where $i=1,2$ denotes the chain index and $s\in[0,L]$ is the contour-length parameter. The chain flexibility is characterized by $l_{\rm{p}}$, which enters the bending potential energy
\begin{equation}\label{H_bending}
    \mathscr{H}_{\rm{B},i}=\frac{\kappa}{2}\int_{0}^{L}ds\left(\frac{\partial\bm{t}_i(s)}{\partial s}\right)^2,
\end{equation}
where $\kappa=k_{\rm{B}}Tl_{p}$ is the bending stiffness in three dimensions \cite{Broedersz,Kleinert}, and $\bm{t}_i(s)=\partial\bm{r}_i(s)/\partial s$ is the unit tangent vector ($\lvert\bm{t}_i(s)\rvert=\bm{1}$) under the assumption that the chain is locally inextensible along its contour. One endpoint of each chain is fixed at the origin [$\bm{r}_1(0)=\bm{r}_2(0)=\bm{0}$], and the two chains are coupled by a single cross-link at $s=L$. The cross-link is modeled as a harmonic potential, $\mathscr{U}=\xi[\bm{r}_1(L)-\bm{r}_2(L)]^2/2$, as defined in the cFJL model. In terms of the tangent vectors, the cross-linking potential energy can be expressed in terms of the tangent vectors as
\begin{equation}\label{H_cross-linking}
    \mathscr{H}_{\rm{c}}=\frac{\xi}{2}\left(\int_{0}^{L}ds\left[\bm{t}_{1}(s)-\bm{t}_{2}(s)\right]\right)^2.
\end{equation}
When each chain of the cWLL is subjected to a pair of tensile forces $\pm f\hat{x}$ applied at its ends, the stretching potential energy, expressed in terms of the tangent vectors, takes the form
\begin{equation}\label{H_stretching}
    \mathscr{H}_{\rm{s}}=-f\int_{0}^{L}ds\left[t_{1,x}(s)+t_{2,x}(s)\right],
\end{equation}
where $t_{i,x}$ denotes the $x$-component of $\bm{t}_{i}$. Collecting the three types of potential energy, the semiflexible system is described by the effective Hamiltonian
\begin{equation}\label{H_cWLL}
    \mathscr{H}_{\rm{cWLL}}=\mathscr{H}_{\rm{B},1}+\mathscr{H}_{\rm{B},2}+\mathscr{H}_{\rm{c}}+\mathscr{H}_{\rm{s}}.
\end{equation}

In the weakly bending regime, where the tangent vectors are nearly aligned with the force axis, the local inextensibility constraint ($\lvert\bm{t}_{i}\rvert=1$) provides a second-order approximation for the longitudinal component in terms of the transverse components [cf.~Eq.~\eqref{b_ij,x}]:
\begin{equation}\label{t_i,x}
    t_{i,x}\approx1-\frac{{\bm{t}_{i,\perp}}^2}{2},\quad\lvert\bm{t}_{i,\perp}\rvert \ll 1,
\end{equation}
where $\bm{t}_{i,\perp}\equiv(t_{i,y},t_{i,z})$ denotes the two-component transverse tangent vector. Substituting Eq.~\eqref{t_i,x} into Eq.~\eqref{H_cWLL} and keeping terms up to quadratic order, we obtain the effective Hamiltonian in the weakly bending regime:
\begin{align}\label{H_cWLL'}
    \mathscr{H}_{\rm{cWLL}}&=\frac{\kappa}{2}\int_{0}^{L}ds\left[\left(\frac{\partial\bm{t}_{1,\perp}(s)}{\partial s}\right)^2+\left(\frac{\partial\bm{t}_{2,\perp}(s)}{\partial s}\right)^2\right]\nonumber\\
    &\quad+\frac{\xi}{2}\left(\int_{0}^{L}ds\left[\bm{t}_{1,\perp}(s)-\bm{t}_{2,\perp}(s)\right]\right)^2-L(f_1+f_2)\nonumber\\
    &\quad+\frac{1}{2}\int_{0}^{L}ds\left[f_1{\bm{t}_{1,\perp}(s)}^2+f_2{\bm{t}_{2,\perp}(s)}^2\right].
\end{align}
As in the cFJL model, we label the applied forces on the two chains as $f_i$ to facilitate later calculations. We point out once again that the present formulation captures only transverse coupling, with longitudinal contributions lying beyond the scope of the current approximation.

\subsection{Partition function}

To construct the canonical partition function of the weakly bent cWLL, we analyze a single-component partial Hamiltonian, as in the cFJL case [Eq.~\eqref{H_cFJL*}]:
\begin{align}\label{H_cWLL*}
      \mathscr{H}_{\rm{cWLL}}^{*}&\equiv\frac{\kappa}{2}\int_{0}^{L}ds\left[\left(\frac{\partial t_{1,\perp}(s)}{\partial s}\right)^2+\left(\frac{\partial t_{2,\perp}(s)}{\partial s}\right)^2\right]\nonumber\\
      &\quad+\frac{\xi}{2}\left(\int_{0}^{L}ds\left[t_{1,\perp}(s)-t_{2,\perp}(s)\right]\right)^2\nonumber\\
      &\quad+\frac{1}{2}\int_{0}^{L}ds\left[f_1{t_{1,\perp}(s)}^2+f_2{t_{2,\perp}(s)}^2\right].
\end{align}
The corresponding partition function is then given by the path integral
\begin{equation}\label{Z_cWLL*}
    \mathscr{Z}_{\rm{cWLL}}^{*}=\int\mathscr{D}[t_{1,\perp}(s)]\int\mathscr{D}[t_{2,\perp}(s)]\exp\left(-\frac{\mathscr{H}_{\rm{cWLL}}^{*}}{k_{\rm{B}}T}\right),
\end{equation}
where $\int\mathscr{D} [t_{i,\perp}(s)]$ denotes the sum over all possible configurations of $t_{i,\perp}(s)$, with $s\in[0,L]$. This expression can be reformulated into a more tractable form by incorporating the relevant boundary conditions. In the physical context of elongational stretching without any lateral forces, the bending moments at the endpoints vanish. Mathematically, this corresponds to vanishing curvature at the chain ends:
\begin{equation}\label{bc}
    \left.\frac{\partial t_{i,\perp}(s)}{\partial s}\right|_{s=0}=\left.\frac{\partial t_{i,\perp}(s)}{\partial s}\right|_{s=L}=0.
\end{equation}

In accordance with these boundary conditions, $t_{i,\perp}(s)$ can be represented as a truncated Fourier cosine series
\begin{equation}\label{t_i}
        t_{i,\perp}(s)=\frac{c_{i0}}{2}+\sum_{m=1}^{M}c_{im}\cos(k_{m}s),
\end{equation}
where $k_m\equiv m\pi/L$. The truncation at a finite number of Fourier modes, $M$, accounts for the coarse-grained nature of the semiflexible polymer model \cite{Alice}. Substituting Eq.~\eqref{t_i} into Eq.~\eqref{H_cWLL*} and applying the orthogonality of the cosine functions, we obtain the Fourier-space representation of the Hamiltonian as
\begin{equation}\label{H_cWLL*_Fourier}
    \frac{\mathscr{H}_{\rm{cWLL}}^{*}}{k_{\rm{B}}T}=A+B,
\end{equation}
where
\begin{equation*}
    A\equiv\frac{L}{8k_{\rm{B}}T}\Big[f_1{c_{10}}^2+f_2{c_{20}}^2+\xi L\left(c_{10}-c_{20}\right)^2\Big],
\end{equation*}
\begin{equation*}
    B\equiv\frac{L}{4k_{\rm{B}}T}\sum_{m=1}^{M}\Big[\left(\kappa{k_m}^2+f_1\right){c_{1m}}^2+\left(\kappa{k_m}^2+f_2\right){c_{2m}}^2\Big].
\end{equation*}
Accordingly, the path integral in Eq.~\eqref{Z_cWLL*} transforms into a (2+2$M$)-dimensional integral in Fourier space, spanning all possible values of the Fourier coefficients $c_{im}$: 
\begin{equation}\label{Z_cWLL*_Fourier}
    \mathscr{Z}_{\rm{cWLL}}^{*}=\mathscr{Z}_{A}\mathscr{Z}_{B},
\end{equation}
where
\begin{equation*}
    \mathscr{Z}_{A}\equiv\int dc_{10}\int dc_{20}\exp\left(-A\right),
\end{equation*}
\begin{equation*}
    \mathscr{Z}_{B}\equiv\int dc_{11}\cdots\int dc_{1M}\int dc_{21}\cdots\int dc_{2M}\exp\left(-B\right).
\end{equation*}
These Gaussian integrals can be evaluated straightforwardly, yielding
\begin{equation*}
    \mathscr{Z}_{A}=\sqrt{\frac{64\pi^2{k_{\rm{B}}}^2T^2}{L^2f_1f_2+\xi L^3(f_1+f_2)}},
\end{equation*}
\begin{equation*}
    \mathscr{Z}_{B}=\prod_{m=1}^{M}\frac{4k_{\rm{B}}TL/\pi\kappa}{\sqrt{(m^2+{\alpha_1}^2)(m^2+{\alpha_2}^2)}},
\end{equation*}
where $\alpha_i\equiv\sqrt{L^2f_i/\pi^2\kappa}$. Finally, the total partition function is assembled as 
\begin{align}\label{Z_cWLL'}
    \mathscr{Z}_{\rm{cWLL}}&=\exp\left(\frac{L(f_1+f_2)}{k_{\rm{B}}T}\right){(\mathscr{Z}_{\rm{cWLL}}^{*})}^2\nonumber\\
    &=\exp\left(\frac{L(f_1+f_2)}{k_{\rm{B}}T}\right)\frac{64\pi^2{k_{\rm{B}}}^2T^2}{L^2f_1f_2+\xi L^3(f_1+f_2)}\nonumber\\
    &\quad\times\prod_{m=1}^{M}\frac{(4k_{\rm{B}}TL/\pi\kappa)^2}{(m^2+{\alpha_1}^2)(m^2+{\alpha_2}^2)}.
\end{align}
In terms of a single force parameter, the partition function reduces to
\begin{equation}\label{Z_cWLL}
    \mathscr{Z}_{\rm{cWLL}}=\frac{\left[8\pi k_{\rm{B}}T\exp(Lf/k_{\rm{B}}T)\right]^2}{L^2f^2+2\xi L^3f}\prod_{m=1}^{M}\left(\frac{4k_{\rm{B}}TL/\pi\kappa}{m^2+\alpha^2}\right)^2,
\end{equation}
where $\alpha\equiv\sqrt{L^2f/\pi^2\kappa}$. As in the previous section, we use the general form only for calculations involving force derivatives. Note that when $\xi=0$, the partition function takes the form $\mathscr{Z}_{\rm{cWLL}}\vert_{\xi=0}=(\mathscr{Z}_{\rm{WLC}})^2$, where  
\begin{equation}\label{Z_WLC}
    \mathscr{Z}_{\rm{WLC}}=\frac{8\pi\exp(Lf/k_{\rm{B}}T)}{Lf/k_{\rm{B}}T}\prod_{m=1}^{M}\frac{4k_{\rm{B}}TL/\pi\kappa}{m^2+\alpha^2}
\end{equation}
is the partition function of a single wormlike chain (WLC) under tension $f$ in the weakly bending regime [Eq.~\eqref{app:Z_WLC}].

\subsection{Average extension and fluctuations}

Although the partition function of the cWLL cannot be cast in closed form, the average projection length along the force axis is analytically accessible. The average extension of the system is obtained from the force derivative of $k_{\rm{B}}T\ln\mathscr{Z}_{\rm{cWLL}}$. Here, the logarithmic form converts the product into a sum, yielding
\begin{equation}\label{X-f_cWLL}
    \langle X\rangle_{\rm{cWLL}}=L-\frac{k_{\rm{B}}T(f+\xi L)}{f^2+2\xi Lf}-\sum_{m=1}^{M}\frac{k_{\rm{B}}TL^2/\pi^2\kappa}{m^2+\alpha^2}.
\end{equation}
For sufficiently large $M$, corresponding to finer resolution of the polymer, the summation can be replaced by the infinite series, as terms with $m>M$ are negligible. Applying the series expansion of the $\coth$ function,
\begin{equation*}
    \sum_{m=1}^{\infty}\frac{1}{m^2+\alpha^2}=\frac{\pi\alpha\coth(\pi\alpha)-1}{2\alpha^2},
\end{equation*}
Eq.~\eqref{X-f_cWLL} is recast in closed form as
\begin{equation}\label{X-f_cWLL_closed}
   \langle X\rangle_{\rm{cWLL}}=\langle X\rangle_{\rm{WLC}}+\langle\Xi\rangle_{\rm{cWLL}},
\end{equation}
where
\begin{equation}\label{X-f_WLC_closed}
    \langle X\rangle_{\rm{WLC}}=L-\frac{k_{\rm{B}}TL}{\sqrt{4\kappa f}}\coth\left(\sqrt{\frac{L^2f}{\kappa}}\right)-\frac{k_{\rm{B}}T}{2f}
\end{equation}
amounts to the average extension of a single wormlike chain in the weakly bending regime [Eq.~\eqref{app:X-f_WLC_closed}], and
\begin{equation}\label{correction_cWLL}
    \langle\Xi\rangle_{\rm{cWLL}}=\frac{k_{\rm{B}}T}{f}-\frac{k_{\rm{B}}T(f+\xi L)}{f^2+2\xi Lf}=\frac{k_{\rm{B}}T\xi L}{f^2+2\xi Lf}
\end{equation}
quantifies the correction due to the cross-linking interaction in the cWLL model. Not surprisingly, the correction term vanishes when $\xi=0$. Interestingly, this expression takes the same functional form as the correction term in the cFJL model [Eq.~\eqref{correction_cFJL}], with $L=Nb$.

The fluctuations of the cWLL are defined and evaluated in the same manner as in the cFJL model [Eqs.~\eqref{fluctuation_l_cFJL} and~\eqref{fluctuation_t_cFJL}]. The longitudinal fluctuations are given by
\begin{align}\label{fluctuation_l_cWLL}
    &\left\langle\left(R_{\parallel}-\langle R_{\parallel}\rangle\right)^2\right\rangle_{\rm{cWLL}}=\frac{{k_{\rm{B}}}^2T^2L^2}{2\kappa f}\csch^2\left(\sqrt{\frac{L^2f}{\kappa}}\right)\nonumber\\
    &\quad+\frac{{k_{\rm{B}}}^2T^2L}{\sqrt{4\kappa f^3}}\coth\left(\sqrt{\frac{L^2f}{\kappa}}\right)-\frac{{k_{\rm{B}}}^2T^2}{f^2}+\frac{2{k_{\rm{B}}}^2T^2}{f^2+2\xi Lf},
\end{align}
and the transverse fluctuations, a more relevant observable to our discussion, are give by
\begin{equation}\label{fluctuation_t_cWLL}
    \langle{\bm{R}_{\perp}}^2\rangle_{\rm{cWLL}}=\frac{4k_{\rm{B}}TL}{f+2\xi L}.
\end{equation}
When $\xi=0$, these fluctuations become twice those of a single wormlike chain [Eqs.~\eqref{fluctuation_l_WLC} and~\eqref{fluctuation_t_WLC}], consistent with the behavior observed in the cFJL model. Moreover, a closer comparison of the transverse fluctuations across the models reveals that they share the same functional form, despite their fundamentally different microscopic descriptions (discrete vs continuous, exclusion vs inclusion of bending elasticity). This echoes the similarity in their cross-link correction terms [Eqs.~\eqref{correction_cFJL} and~\eqref{correction_cWLL}], both of which originate exclusively from the transverse coupling.

Owing to this shared functional structure of the transverse fluctuations, the key characteristics established in the cFJL model naturally carry over to the present case. The presence of the cross-link suppresses the transverse fluctuations, and the suppression becomes more pronounced as the system size increases. In the absence of the cross-link, the transverse fluctuations grow linearly with $L$, whereas in its presence they approach a constant value, $2k_{\rm{B}}T/\xi$. In the thermodynamic limit, the relative magnitude of the transverse fluctuations, $\sqrt{\langle{\bm{R}_{\perp}}^2\rangle}/\langle X\rangle$, scales as $L^{-1/2}$ in the uncoupled case, whereas it scales as $L^{-1}$ in the coupled case. This demonstrates that the single cross-link effectively reduces the spatial separation between the chain ends, thereby supporting, here as well, the identification of the cWLL as a looped configuration.

In the wormlike chain model, the emergence of hyperbolic functions with argument $\sqrt{L^2f/\kappa}$ motivates the introduction of a new length scale, $l_{\rm{m}}\equiv\sqrt{\kappa/f}$. We refer to $l_{\rm{m}}$ as the (orientational) memory length, following Ref.~\cite{Alice}. This memory length, also known as the deflection length, offers an alternative measure of the chain's rigidity under nematic confinement \cite{Odijk,deGennes2}. In this context, strong stretching imposes the nematic (orientational) confinement, with the force axis acting as the nematic director. The introduction of the memory length delineates two important limiting regimes, $l_{\rm{m}}\ll L$ and $l_{\rm{m}}\gg L$, each providing further insight into the tensile elasticity.

\subsubsection{Ordinary cWLL in the regime \texorpdfstring{$l_{\rm{m}}\ll L$}{lm ≪ L}}\label{sec:ordinary}

The case $l_{\rm{m}}\ll L$ characterizes the weakly bending regime imposed by strong stretching ($Lf/k_{\rm{B}}T\gg l_{\rm{p}}/L$). This case further assumes $l_{\rm m}\ll l_{\rm p}$, ensuring consistency with the weakly bending description. As long as these two conditions are met, the relative magnitude between $l_{\rm{p}}$ and $L$ is not essential. Nonetheless, our focus here is on ordinary semiflexible polymers whose persistence length is less than or comparable to the total contour length ($l_{\rm{p}}\lesssim L$), for which strong stretching---rather than a large persistence length---is primarily responsible for the weakly bending behavior. In this regime, where $\coth(L/l_{\rm{m}})\approx1$ and $\csch(L/l_{\rm{m}})\approx0$, the average extension [Eq.~\eqref{X-f_cWLL_closed}] becomes
\begin{equation}\label{X-f_cWLL_ordinary}
    \langle X\rangle_{\rm{cWLL}}\approx L-\frac{k_{\rm{B}}TL}{\sqrt{4\kappa f}}-\frac{k_{\rm{B}}T}{2f}+\frac{k_{\rm{B}}T\xi L}{f^2+2\xi Lf},
\end{equation}
and the longitudinal fluctuations [Eq.~\eqref{fluctuation_l_cWLL}] simplify to
\begin{equation}\label{fluctuation_l_cWLL_ordinary}
    \left\langle\left(R_{\parallel}-\langle R_{\parallel}\rangle\right)^2\right\rangle_{\rm{cWLL}}\approx\frac{{k_{\rm{B}}}^2T^2L}{\sqrt{4\kappa f^3}}-\frac{{k_{\rm{B}}}^2T^2}{f^2}+\frac{2{k_{\rm{B}}}^2T^2}{f^2+2\xi Lf}.
\end{equation}
To leading order, the force--extension relation exhibits the characteristic $f^{-1/2}$ dependence of a wormlike chain under strong stretching \cite{Marko-Siggia}, and the ratio of the longitudinal fluctuations (i.e., the mismatch of the two chain ends), $\sqrt{\big\langle\left(R_{\parallel}-\langle R_{\parallel}\rangle\right)^2\big\rangle}$, to the entropic shrinkage $\langle X\rangle-L$ scales as $f^{-1/4}$, vanishing asymptotically. Thus, the two chains effectively behave as if they fluctuate longitudinally in concert, further validating the loop interpretation in this strong stretching regime.

In parallel with the cFJL analysis, we revisit the tensile elasticity of the cWLL from the angle of a ``single-loop'' structure and contrast it with a single-chain counterpart. Consider the cWLL and the WLC in the regime $l_{m}\ll L$, under the same total stretching force $f$. For the loop, the leading-order average extension is given by
\begin{equation}\label{X-f_cWLL_ordinary_approx}
    \langle X\rangle_{\rm{cWLL}}=L\left(1-\frac{k_{\rm{B}}T}{\sqrt{2\kappa f}}\right),
\end{equation}
whereas for the single chain [Eq.~\eqref{X-f_WLC_ordinary}], the leading-order behavior is
\begin{equation}\label{X-f_WLC_ordinary_approx}
    \langle X\rangle_{\rm{WLC}}=L\left(1-\frac{k_{\rm{B}}T}{\sqrt{4\kappa f}}\right).
\end{equation}
Following the approach taken for the cFJL [cf.~Eq.~\eqref{compliance_FJ}], we obtain the tensile compliance in the linear-response regime as
\begin{equation}\label{compliance_W_ordinary}
    \frac{\delta X_{\rm{cWLL}}}{\delta f}=\frac{k_{\rm{B}}TL}{\sqrt{8\kappa {f_0}^3}},\quad\frac{\delta X_{\rm{WLC}}}{\delta f}=\frac{k_{\rm{B}}TL}{\sqrt{2\kappa{f_0}^3}}.
\end{equation}
Consistent with the freely jointed model, the compliance of the loop is half that of the single chain, making the loop effectively twice as stiff.

The appearance of this shared twofold difference in tensile elasticity across the models reflects an underlying connection, which is illuminated through the concept of memory length. In terms of $l_{\rm{m}}$, the force--extension relations in Eqs.~\eqref{X-f_WLC_ordinary_approx} and~\eqref{X-f_cWLL_ordinary_approx} become
\begin{equation}\label{X-f_WLC_ordinary_memory}
    \langle X\rangle_{\rm{WLC}}=L\left(1-\frac{k_{\rm{B}}T}{2\left[l_{\rm{m}}(f)\right]f}\right),
\end{equation}
\begin{equation}\label{X-f_cWLL_ordinary_memory}
    \langle X\rangle_{\rm{cWLL}}=L\left(1-\frac{2k_{\rm{B}}T}{2\left[l_{\rm{m}}(f/2)\right]f}\right),
\end{equation}
where $l_{\rm{m}}$ is written explicitly as a function of $f$ to highlight its dependence on the local force, and the factors of $2$ are intentionally left uncanceled in Eq.~\eqref{X-f_cWLL_ordinary_memory}. In these forms, the force--extension relations mirror those of the freely jointed systems [Eqs.~\eqref{X-f_cFJL} and~\eqref{X-f_FJC}], with effective bond lengths $\widetilde{b}_{\rm{WLC}}=2l_{\rm{m}}(f)$ and $\widetilde{b}_{\rm{cWLL}}=2l_{\rm{m}}(f/2)$. Despite the force dependence of the effective bond lengths in the wormlike systems, the twofold difference remains intact, as this dependence automatically adjusts to preserve the ratio. Note that for a free WLC, the effective bond length, also known as the Kuhn segment length, is given by $b_{\rm{K}}=2l_{\rm{p}}$ \cite{Rubinstein,Grosberg}. Under strong tension, the persistence length $l_{\rm{p}}$ gives way to the more appropriate, force-dependent memory length $l_{\rm{m}}(f)$. This correspondence clarifies the meaning of the term ``orientational memory length'': under strong stretching, $l_{\rm{m}}$ serves as an independent segment length scale over which the chain retains, or ``remembers'', its orientation. In this sense, the orientational memory is greater in the loop because the two chains forming the loop experience smaller local forces. These smaller forces cause less disruption to segment orientations and thus allow the chains to maintain directional correlations over longer distances. It is interesting to note that the strong stretching limit can be expressed as $l_{\rm{m}}f\gg k_{\rm{B}}T$ (analogous to the FJC condition $bf\gg k_{\rm{B}}T$), which is equivalent to requiring $l_{\rm{m}}\ll l_{\rm{p}}$.

\subsubsection{Rodlike cWLL in the regime \texorpdfstring{$l_{\rm{m}}\gg L$}{lm ≫ L}}\label{sec:rodlike}

Apart from the strong stretching regime analyzed above, the case $l_{\rm{m}}\gg L$ corresponds to the weakly bending regime dominated by a large persistence length ($l_{\rm{p}}/L\gg Lf/k_{\rm{B}}T$) under a moderately strong tensile force. The force $f$ should be sufficiently small for the chain to retain long-range orientational memory, yet large enough to ensure alignment along the force axis. More specifically, this case applies to rodlike polymers whose persistence length is much greater than the total contour length ($l_{\rm{p}}\gg L$), with $Lf\gg k_{\rm B}T$. These conditions together imply that the three length scales satisfy $L\ll l_{\rm{m}}\ll l_{\rm{p}}$. In this regime, the average extension [Eq.~\eqref{X-f_cWLL_closed}] becomes
\begin{equation}\label{X-f_cWLL_rodlike}
    \langle X\rangle_{\rm{cWLL}}\approx L-\frac{k_{\rm{B}}TL^2}{6\kappa}-\frac{k_{\rm{B}}T}{f}+\frac{k_{\rm{B}}T\xi L}{f^2+2\xi Lf},
\end{equation}
and the longitudinal fluctuations [Eq.~\eqref{fluctuation_l_cWLL}] simplify to
\begin{equation}\label{fluctuation_l_cWLL_rodlike}
    \left\langle\left(R_{\parallel}-\langle R_{\parallel}\rangle\right)^2\right\rangle_{\rm{cWLL}}\approx\frac{{k_{\rm{B}}}^2T^2L^4}{72\kappa^2}+\frac{2{k_{\rm{B}}}^2T^2}{f^2+2\xi Lf}.
\end{equation}
Both expressions are obtained by applying the Taylor series expansions of the $\coth$ and $\csch$ functions to second leading order (the lowest order that gives rise to a nontrivial result). In contrast to the previous regime, the force--extension relation exhibits an $f^{-1}$ dependence at leading order, characteristic of a freely jointed chain under strong stretching, and the ratio of the longitudinal fluctuations $\sqrt{\big\langle\left(R_{\parallel}-\langle R_{\parallel}\rangle\right)^2\big\rangle}$ to the entropic shrinkage $\langle X\rangle-L$ is of order unity. Thus, although the transverse mismatch---controlled through $\xi$ as shown in Eq.~\eqref{fluctuation_t_cWLL}---can be made negligibly small, the two chains nevertheless fluctuate longitudinally in an essentially independent fashion, so the configuration cannot be regarded as a \textit{bona fide} loop in the same sense as in the previous case.

We once again treat the cWLL as an effective single loop and compare it with a single WLC. Focusing on the leading-order behavior under the same total stretching force $f$, the average extensions of the loop and the single chain [Eq.~\eqref{X-f_WLC_rodlike}] are given by
\begin{equation}\label{X-f_cWLL_rodlike_approx}
    \langle X\rangle_{\rm{cWLL}}=L\left(1-\frac{k_{\rm{B}}TL}{6\kappa}\right)-\frac{2k_{\rm{B}}T}{f},
\end{equation}
\begin{equation}\label{X-f_WLC_rodlike_approx}
    \langle X\rangle_{\rm{WLC}}=L\left(1-\frac{k_{\rm{B}}TL}{6\kappa}\right)-\frac{k_{\rm{B}}T}{f}.
\end{equation}
In this regime as well, the loop is effectively twice as stiff as the single chain. Here, the memory length does not emerge naturally; the persistence length alone suffices to characterize the system. Rewriting the force--extension relations in the following form highlights their rodlike behavior more clearly:
\begin{equation}\label{X-f_W_rodlike}
    \langle X\rangle_{\rm{cWLL}}=\widetilde{L}\left(1-\frac{2k_{\rm{B}}T}{\widetilde{L}f}\right),\ \langle X\rangle_{\rm{WLC}}=\widetilde{L}\left(1-\frac{k_{\rm{B}}T}{\widetilde{L}f}\right),
\end{equation}
where $\widetilde{L}\equiv L(1-L/l_{\rm{p}})$. The rodlike WLC can be viewed as a single rigid rod of length $\widetilde{L}$, the extreme case of a freely jointed chain with $N=1$ and $b=\widetilde{L}$. In this rodlike regime, the cWLL behaves as two parallel such rods, leading to the factor of two in the entropic correction.

\section{Polymer Necklace}\label{sec4}

\begin{figure}
    \includegraphics[width=8.6cm]{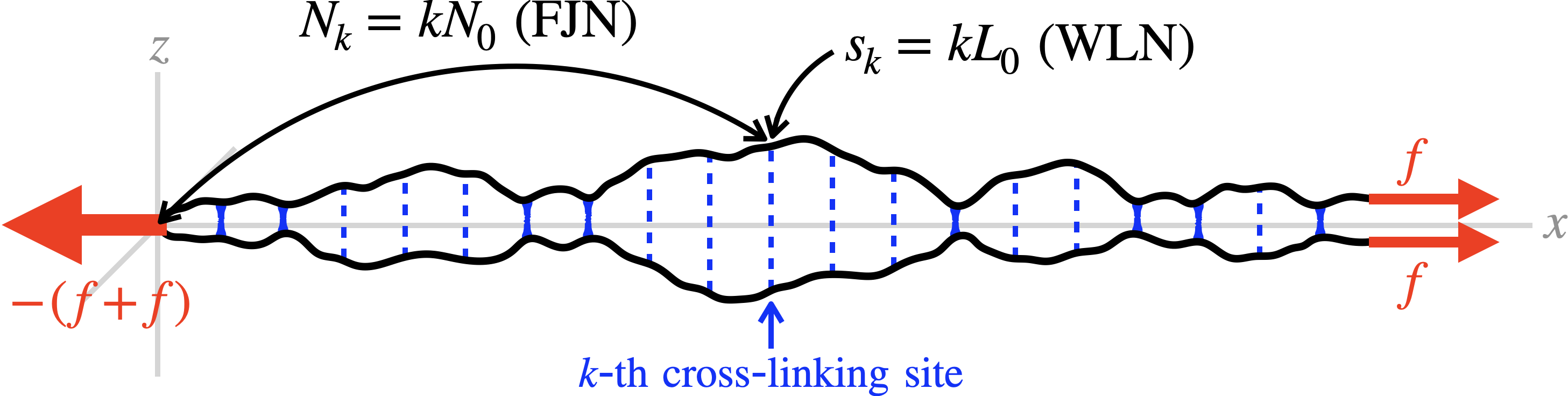}
    \caption{Schematic diagram of the polymer necklace under strong tension. The two chains are interconnected by multiple reversible cross-links, each of which can be either bound (blue solid line) or unbound (blue dotted line). Cross-linking sites are regularly spaced along the contour at positions $N_k=kN_0$ for the freely jointed necklace (FJN) and $s_k=kL_0$ for the wormlike necklace (WLN), where $N_0$ and $L_0$ are the respective interval parameters for each model.}
    \label{fig:PN}
\end{figure}

The polymer necklace is conceived as an extension of the cFJL or cWLL models, in which multiple reversible cross-links are regularly distributed between the chain pair to form a necklace-like system (see Fig.~\ref{fig:PN}). Specifically, we consider an infinitely long pair of freely jointed or wormlike chains with annealed disorder in the local cross-linking states, where each cross-linking site can be either bound or unbound. In both models, the necklace is in the strong stretching (and thus weakly bending) regime so that the constituent chains are directed along the force axis. Our main goal is to investigate the binding behavior of this directed-polymer necklace, with the primary observable being the mean fraction of the bound cross-links in the thermodynamic limit.

\begin{figure*}
    \includegraphics[width=17cm]{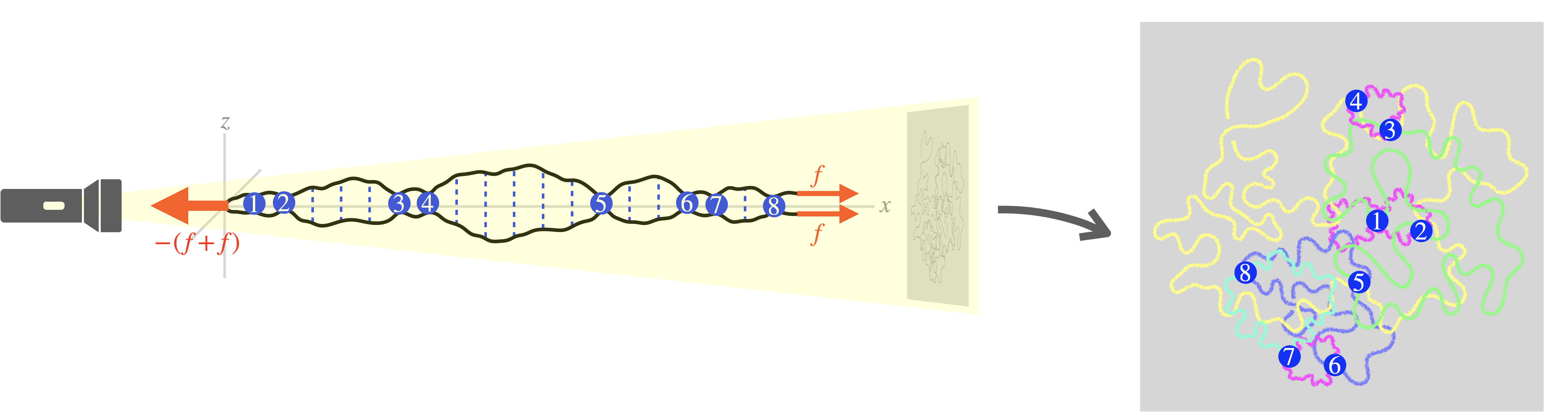}
    \caption{Schematic illustration of the Gaussian slinky representation, highlighting the statistical equivalence between cross-linking events in the strongly stretched necklace (left) and loop-formation events in its transverse-plane projection (the Gaussian slinky, right). The matched numbering indices are included in the two panels to emphasize the counting equivalence, and distinct colors are used in the Gaussian slinky to represent different loop sizes for visual distinguishability.}
    \label{fig:GS}
\end{figure*}

\subsection{Freely jointed necklace}\label{sec:FJN}

The freely jointed necklace (FJN) consists of a pair of identical freely jointed chains, with cross-linking sites equally spaced at positions $N_k=kN_0$ ($k=1,2,3,\cdots,\nu$), as illustrated in Fig.~\ref{fig:PN}. Each site can transition between bound and unbound states, making the cross-linking reversible. Under symmetric strong stretching, the effective Hamiltonian [cf.~Eq.~\eqref{H_cFJL'}] is given by
\begin{align}\label{H_FJN}
    \mathscr{H}_{\rm{FJN}}&=\frac{\xi}{2}\sum_{k=1}^{\nu}\theta_k\left(\sum_{j=1}^{kN_{0}}(\bm{b}_{1j,\perp}-\bm{b}_{2j,\perp})\right)^2-\sum_{k=1}^{\nu}\theta_k\mu\nonumber\\
    &\quad-2\nu N_0 bf+\frac{f}{2b}\sum_{j=1}^{\nu N_0}\left({\bm{b}_{1j,\perp}}^2+{\bm{b}_{2j,\perp}}^2\right),
\end{align}
where $\theta_{k}$ is a binary variable indicating whether the $k$th site is in the bound ($\theta_k\!=\!1$) or unbound ($\theta_k\!=\!0$) state. The fluctuations of local occupation numbers are governed by the energy parameter $\mu$, which can be interpreted as a chemical potential in the grand canonical ensemble, or as an activation energy assuming an Arrhenius-type behavior. In principle, the mean fraction of bound cross-links can be obtained by constructing the partition function of the FJN and differentiating its logarithm with respect to, this time, $\mu$. In practice, however, this direct approach using a technique similar to that of the previous sections is not analysis-friendly. Fortunately, this motivates an alternative route that leverages insights from the transverse fluctuations.

In the strong stretching regime, the transverse fluctuations of the FJC are given by $\langle{\bm{R}_{\perp}}^2\rangle_{\rm{FJC}}=2Nk_{\rm{B}}Tb/f$. This implies that the projected ``trajectory'' (polymer conformation) onto the plane normal to the force axis follows the statistics of a two-dimensional random walk with $N$ steps and an effective step size of $\ell=\sqrt{2k_{\rm{B}}Tb/f}$. Note that $\ell$ is simply the root-mean-square transverse excursion of an individual segment, $\sqrt{\langle{\bm{b}_{j,\perp}}^2\rangle}$. For $N\gg1$, the trajectory becomes Gaussian and can be modeled as a Gaussian chain characterized by the probability distribution 
\begin{equation}\label{rho}
    \rho(N,\bm{R}_{\perp})=\frac{1}{\pi N{\ell}^2}\exp\left(-\frac{{\bm{R}_{\perp}}^2}{N{\ell}^2}\right).
\end{equation}
Thus, if the FJN is projected onto the transverse plane, it appears as a spatial superposition (or planar stack) of many Gaussian loops, each with a size (degree of polymerization) proportional to the basic unit $N_0\gg1$, as illustrated in Fig.~\ref{fig:GS}. We refer to this projected conformation of the necklace---namely, its shadow in the transverse plane---as the \textit{Gaussian slinky}.

This two-dimensional representation provides a useful mapping between key physical observables of the necklace and geometric features of the slinky. In the necklace, the total number of cross-links (more precisely, the total number of cross-linking sites) is denoted by $\nu$, and the number of bound cross-links currently engaged (i.e., the occupied cross-linking sites) is denoted by $\nu_{\rm{c}}$. In the slinky, these quantities correspond to the maximum number of loops (when all cross-links are bound) and the number of loops currently present, respectively. Therefore, the mean fraction of bound cross-links in the three-dimensional model, $\langle\nu_{\rm{c}}/\nu\rangle$, can be inferred from the fraction of realized loops in the two-dimensional picture. An important advantage of the Gaussian slinky model is that the sequential loops are structurally identical and independent. This facilitates the use of a powerful tool for analyzing phase behavior in the thermodynamic limit: the generating function method \cite{Fisher2,Grosberg,Rudnick2,NGH3}.

As a first step, we evaluate the partition function of the Gaussian loop. Consider two identical Gaussian chains of degree of polymerization $kN_0$ ($k\in\mathbb{N}$) and effective bond length $\ell=\sqrt{2k_{\rm{B}}Tb/f}$, each embedded in two-dimensional space with one endpoint fixed at the origin. Their end-to-end vectors are denoted by $\bm{R}_{1,\perp}$ and $\bm{R}_{2,\perp}$, respectively. Note that although the force $f$ does not act directly on the chains, it influences the free Gaussian chain model through the projected parameter $\ell(f)$. The loop conformation can be enforced by introducing a confining potential well over a small area $a\ll N_0\,{\ell}^2$, within which $\bm{R}_{1,\perp}$ and $\bm{R}_{2,\perp}$ are constrained to coincide \cite{Grosberg}. The condition $a\ll N_0\,{\ell}^2$ ensures that loop closure is confined to a region much smaller than the natural wandering scale of the chains, $\langle{\bm{R}_{i,\perp}}^2\rangle$, so that the conformation can be regarded as a well-defined loop. If $a$ were comparable to the fluctuation scale, the distinction between looped and unlooped states would disappear, and thus the necklace model of alternating two-type sequences, together with its Gaussian slinky representation, would break down. With this smallness condition satisfied, the looping probability can be derived using a delta-function constraint, 
\begin{align}\label{P_l}
    P_{\mathcal{O}}&=\int d^2R_{1,\perp}\int d^2R_{2,\perp}\,\rho(kN_0,\bm{R}_{1,\perp})\,\rho(kN_0,\bm{R}_{2,\perp})\nonumber\\
    &\quad\times a\,\delta(\bm{R}_{1,\perp}-\bm{R}_{2,\perp})=\frac{a}{2\pi{\ell}^2kN_0}.
\end{align}
Since the partition function acts as a statistical weight proportional to the number of corresponding microstates, it can be identified with the probability (up to normalization). Incorporating the potential depth $\varepsilon_a>0$ of the confining well---an energetic contribution not encoded in the geometric probability---the partition function of the loop takes the form
\begin{equation}\label{Z_GL}
    \mathscr{Z}(kN_0)=\frac{1}{2\pi}\left(\frac{a}{{\ell}^2}\right)(kN_0)^{-1}\exp\left(\frac{\varepsilon}{k_{\rm{B}}T}\right),
\end{equation}
where an effective energy parameter $\varepsilon$ emerges from the two-state nature of the loop (looped or unlooped), defined through the relation \cite{NGH2} 
\begin{equation}\label{epsilon}
    \exp\left(\frac{\varepsilon}{k_{\rm{B}}T}\right)=\exp\left(\frac{\varepsilon_a}{k_{\rm{B}}T}\right)-1.
\end{equation}
Physically, $\varepsilon$ plays the same role as $\mu$: $\varepsilon$ controls the statistical weight of the looped state relative to the unlooped state, just as $\mu$ does for cross-link binding. If the confining potential depth is zero ($\varepsilon_a=0$), the loop carries zero statistical weight; in other words, loop formation requires an attractive well.

We now construct the grand partition function of the Gaussian slinky, which serves as the formal generating function for its looping statistics and thus for the cross-linking statistics of the necklace. To this end, we classify the constituent loops of the slinky into two types based on their sizes: type A loops of size $N_0$, and type B loops of larger size $kN_0$ with $k\geq2$. The partial grand partition functions corresponding to each type are obtained by
\begin{equation}\label{G_A}
    \mathscr{G}_{\rm{A}}(q)=\sum_{n=1}^{\infty}\left[\mathscr{Z}(N_0)\right]^nq^n=\frac{wq}{1-wq},
\end{equation}
\begin{equation}\label{G_B}
    \mathscr{G}_{\rm{B}}(q)=\sum_{n=2}^{\infty}\mathscr{Z}(nN_0)q^n=-w\ln(1-q)-wq,
\end{equation}
where $q$ is the fugacity associated with the basic unit (pair or loop) composed of two chains of size $N_0$, and $w\equiv\mathscr{Z}(N_0)$ is the statistical weight of a type A loop. Considering all possible configurations of the slinky in which type A and type B loops occur in alternating sequence (including those that begin and end with either type), the total grand partition function of the Gaussian slinky is constructed as a geometric series, yielding
\begin{equation}\label{G}
    \mathscr{G}(q)=\frac{\mathscr{G}_{\rm{A}}(q)+\mathscr{G}_{\rm{B}}(q)+2\mathscr{G}_{\rm{A}}(q)\mathscr{G}_{\rm{B}}(q)}{1-\mathscr{G}_{\rm{A}}(q)\mathscr{G}_{\rm{B}}(q)}.
\end{equation}

\begin{figure*}
    \sidesubfloat[]{\includegraphics[width=0.4\textwidth]{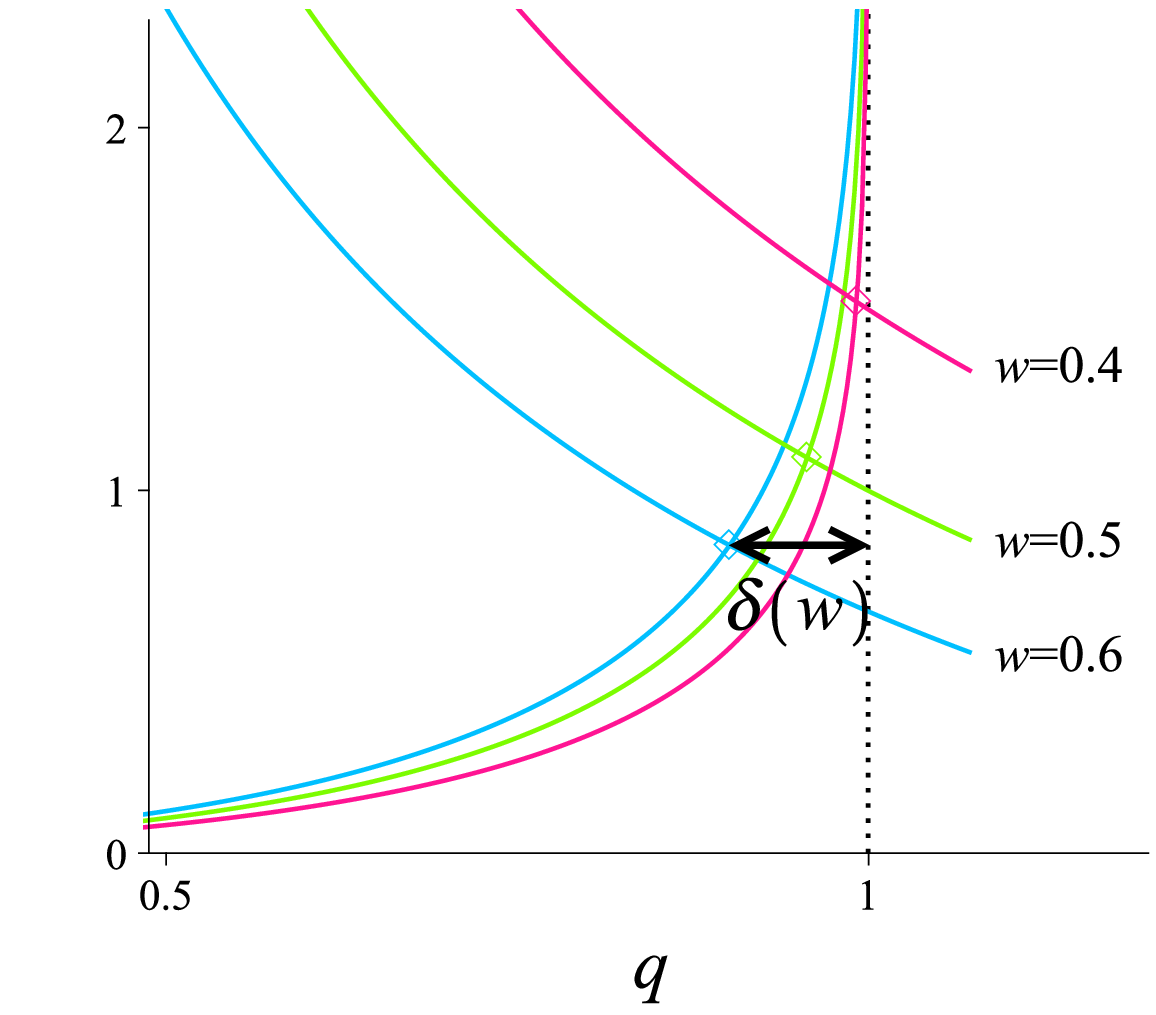}\label{fig:delta_w<1}}
    \hfill
    \sidesubfloat[]{\hspace{5mm}\includegraphics[width=0.4\textwidth]{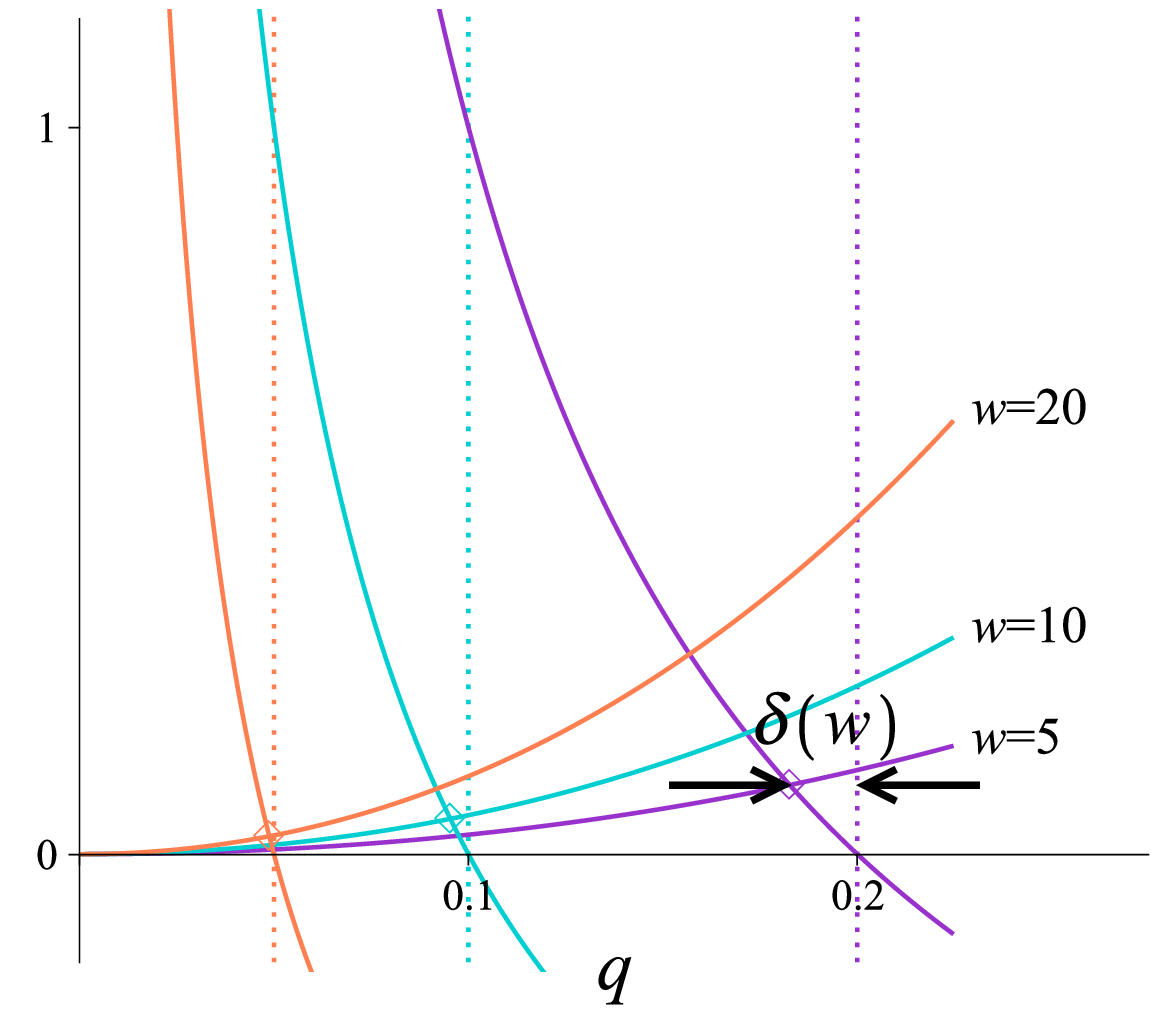}\label{fig:delta_w>1}}
    \caption{Graphical solution of the equation $\mathscr{G}_{\rm{B}}(q)=1/\mathscr{G}_{\rm{A}}(q)$ for different values of the statistical weight $w$. Each increasing curve shows $\mathscr{G}_{\rm{B}}(q)$, and each decreasing curve shows $1/\mathscr{G}_{\rm{A}}(q)$. The intersection points (empty diamonds) give the dominant singularity $q^{*}(w)$ of the total generating function $\mathscr{G}(q)$. (a) For $w<1$, $\delta(w)=1-q^{*}(w)$ measures the deviation from the singularity of $\mathscr{G}_{\rm{B}}(q)$ at $q_{\rm{B}}^{*}=1$ (dotted line). As $w$ decreases, $q^{*}(w)\to1$ with $\delta(w)\ll1$. (b) For $w>1$, $\delta(w)=q_{\rm{A}}^{*}-q^{*}(w)$ measures the deviation from the singularity of $\mathscr{G}_{\rm{A}}(q)$ at $q_{\rm{A}}^{*}=1/w$ (dotted lines). As $w$ increases, $q^{*}(w)\to1/w$ with $\delta(w)\ll1$.}
    \label{fig:q*}
\end{figure*}

In the thermodynamic limit, the phase behavior of the system is imprinted in the smallest positive singularity of $\mathscr{G}(q)$ \cite{Fisher2,NGH3}. Denoting this singularity by $q^{*}$, the limiting free energy density (free energy per basic unit) is given by
\begin{equation}\label{psi}
    \psi=k_{\rm{B}}T\ln q^{*},
\end{equation}
and the mean fraction of realized loops (relative to the maximum possible number, attainable when the slinky consists only of type A loops)---equivalently, the mean fraction of bound cross-links---is obtained via
\begin{equation}\label{n_c}
    \langle n_{\rm{c}}\rangle=\lim_{\nu\to\infty}\left\langle\frac{\nu_{\rm{c}}}{\nu}\right\rangle=-\frac{\partial\psi}{\partial\varepsilon}.
\end{equation}
A graphical analysis shows that $q^{*}$ corresponds to the root of the equation $\mathscr{G}_{\rm{B}}(q)=1/\mathscr{G}_{\rm{A}}(q)$:
\begin{equation}\label{eq_q}
    -w\ln(1-q)-wq=\frac{1}{wq}-1.
\end{equation}
Moreover, the smooth dependence of $q^{*}$ on control parameters such as $T$, $\varepsilon$, and $f$ demonstrates that the system undergoes no phase transition. In fact, the existence of a phase transition and, if so, its order are universal features of the necklace model and are already manifested in Eq.~\eqref{Z_GL}. A power-law dependence of the form $\mathscr{Z}(kN_0)\sim(kN_0)^{-\gamma}$ leads to a phase transition only if the exponent satisfies $\gamma>1$ \cite{Fisher2,Kafri2}. In our case, $\gamma=1$, which confirms the absence of a phase transition.

Having addressed the broader question of phase transitions, we now focus on evaluating $\langle n_{\rm{c}}\rangle$ by determining $q^{*}$. Although Eq.~\eqref{eq_q} does not have a closed-form solution, $q^{*}$ displays a clear asymptotic tendency: it approaches the singularity of $\mathscr{G}_{\rm{B}}$, $q_{\rm{B}}^{*}=1$, as $w\to0$, and that of $\mathscr{G}_{\rm{A}}$, $q_{\rm{A}}^{*}=1/w$, as $w\to\infty$ (see Fig.~\ref{fig:q*}). Based on this behavior, $q^{*}$ can be extracted explicitly in two asymptotic regimes characterized by the magnitude of $w$. To clarify the dependence of $w$ on the underlying physical quantities, we rewrite it by restoring all the relevant control parameters:
\begin{equation}\label{w}
    w=\frac{\sigma}{N_0}\left(\frac{bf}{k_{\rm{B}}T}\right)\exp\left(\frac{\varepsilon}{k_{\rm{B}}T}\right),
\end{equation}
where $\sigma\equiv a/4\pi b^2$ denotes the relative (dimensionless) areal range of the looping interaction. We treat the cases $w<1$ and $w>1$ separately, as they roughly delineate two asymptotic regimes with distinct levels of stretching intensity, both still falling within the strong stretching regime.

\begin{figure*}
    \sidesubfloat[]{\includegraphics[width=0.28\textwidth]{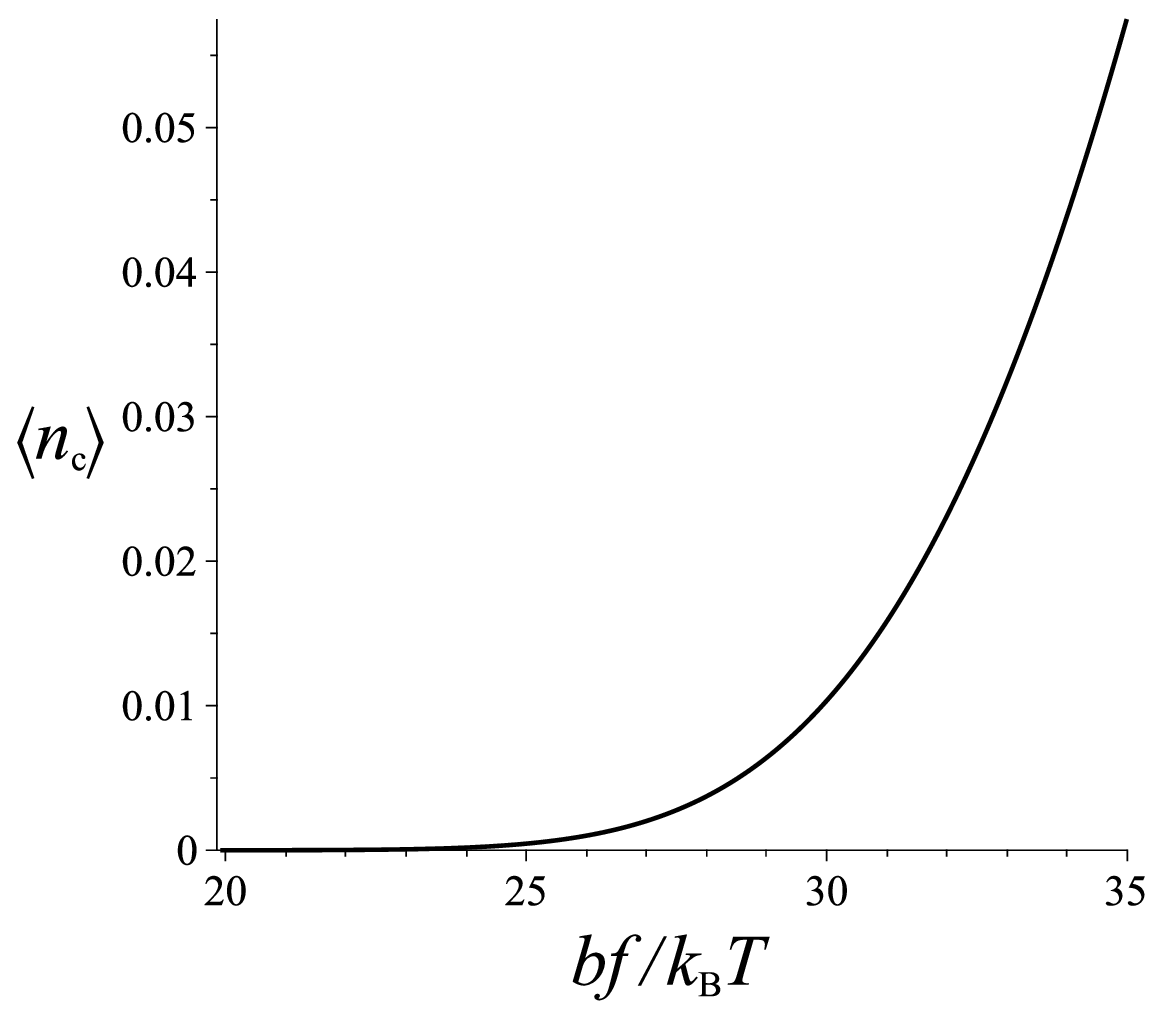}\label{fig:n_c_FJN_w<1}}
    \hfill
    \sidesubfloat[]{\includegraphics[width=0.28\textwidth]{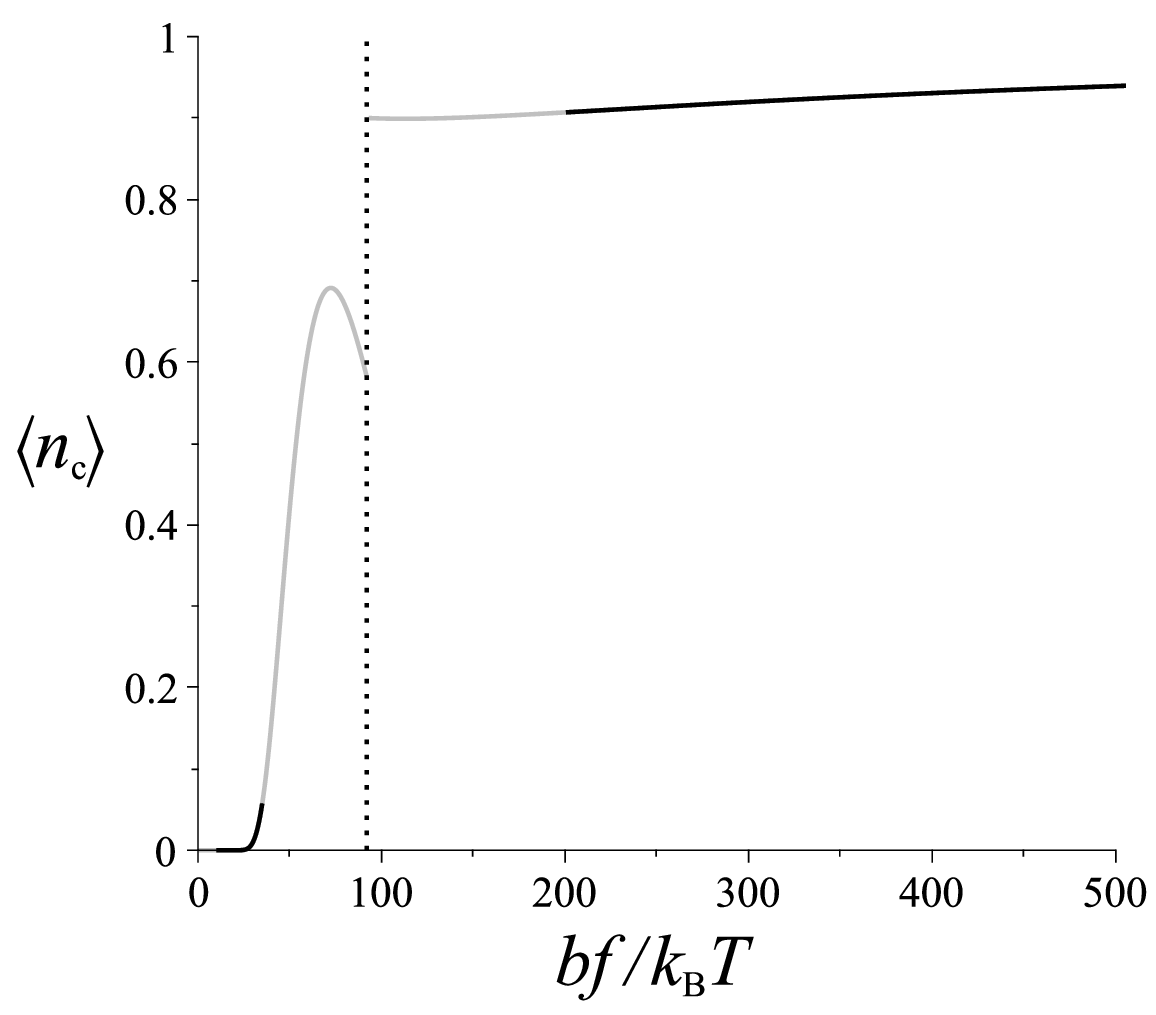}\label{fig:n_c_FJN_overall}}
    \hfill
    \sidesubfloat[]{\includegraphics[width=0.28\textwidth]{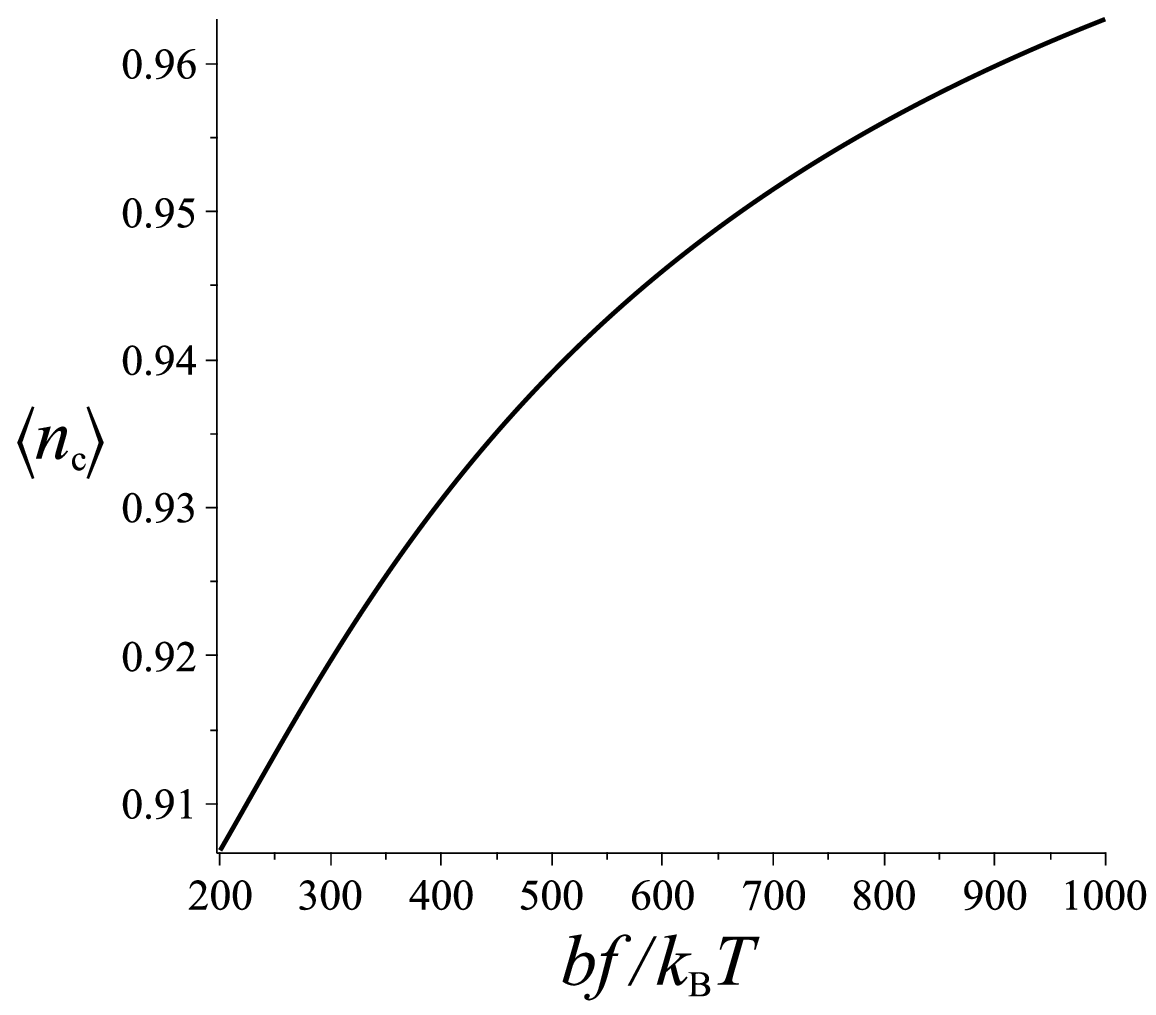}\label{fig:n_c_FJN_w>1}}
    \caption{Mean fraction of bound cross-links $\langle n_{\rm{c}}\rangle$ in the FJN as a function of the dimensionless force $bf/k_{\rm{B}}T$, shown for parameter values $N_0=2025$, $\sigma=0.001$, and $\varepsilon/k_{\rm{B}}T=10$. (a) Weakly bound regime ($w<1$). (b) The vertical dotted line marks the reference point $w=1$, corresponding to $bf/k_{\rm{B}}T\approx92$. The gray segments, which exhibit an unphysical peak, lie outside the domain of validity of the asymptotic approximation and are shown only to illustrate the overall trend. The reliable region (black segments) is determined heuristically to ensure $\delta(w)\ll1$ and should be regarded as the physically meaningful prediction. (c) Strongly bound regime ($w>1$). The model's validity condition, $bf/k_{\rm{B}}T\!\ll\!N_0/2\pi\sigma\approx3.2\!\times\!10^5$, prevents arbitrarily large forces.}
    \label{fig:n_c_FJN}
\end{figure*}

In the regime $w<1$, we define $\delta(w)$ as the distance between $q^{*}$ and $q_{\rm{B}}^{*}=1$, as illustrated in Fig.~\subref*{fig:delta_w<1}. The asymptotic behavior $\delta(w)\ll1$ sets in appreciably early, making it reasonable to adopt $w<1$ as a practical marker for this regime rather than requiring the stricter condition $w\ll1$. Accordingly, we take $w<1$ to indicate that $w$ is sufficiently small for $\delta(w)$ to be asymptotically small. In this regime, we express $q^{*}$ as
\begin{equation}\label{q*_w<1}
    q^{*}=1-\delta(w),\quad\delta(w)\ll1.
\end{equation}
Substituting this expression into Eq.~\eqref{eq_q}, we obtain the asymptotic approximation
\begin{equation}\label{delta_FJN_w<1}
    \delta(w)\approx\exp\left(-\frac{1}{w^2}+\frac{1}{w}-1\right).
\end{equation}
The free energy density is then given by
\begin{equation}\label{psi_FJN_w<1}
    \psi=k_{\rm{B}}T\ln\left[1-\exp\left(-\frac{1}{w^2}+\frac{1}{w}-1\right)\right],
\end{equation}
and, using Eq.~\eqref{n_c}, the mean fraction of bound cross-links is obtained as
\begin{equation}\label{n_c_FJN_w<1}
    \langle n_{\rm{c}}\rangle=\frac{\left(\frac{2}{w^2}-\frac{1}{w}\right)\exp\left(-\frac{1}{w^2}+\frac{1}{w}-1\right)}{1-\exp\left(-\frac{1}{w^2}+\frac{1}{w}-1\right)}.
\end{equation}
To visualize its force dependence, we plot $\langle n_{\rm{c}}\rangle$ as a function of the dimensionless force $bf/k_{\rm{B}}T$ in Fig.~\subref*{fig:n_c_FJN_w<1}. The plotted range covers values of $bf/k_{\rm{B}}T$ on the order of $10$, where the strong stretching behavior of the chain is well-developed (e.g., $\langle X\rangle_{\rm{FJC}}/Nb\approx90\%$ at $bf/k_{\rm{B}}T=10$). It is restricted from above by the condition $w<1$, which corresponds to $bf/k_{\rm{B}}T<N_0/\sigma\exp(\varepsilon/k_{\rm{B}}T)$. The practical cutoff is chosen heuristically to lie safely below this upper limit, ensuring $\delta(w)\ll1$ throughout so that the asymptotic approximation for $\langle n_{\rm{c}}\rangle$ is valid [Fig.~\subref*{fig:n_c_FJN_overall}]. Within the strong stretching regime, Eq.~\eqref{n_c_FJN_w<1} describes the mean fraction of bound cross-links in a directed configuration of the FJN associated with a lower level of stretching intensity, capturing the onset of strong stretching behavior. Reflecting its low binding affinity, we refer to this regime as the weakly bound regime.

In the regime $w>1$, we redefine $\delta(w)$ as the distance between $q^{*}$ and $q_{\rm{A}}^{*}=1/w$, as illustrated in Fig.~\subref*{fig:delta_w>1}. Analogous to the previous case, we take $w>1$ as a practical marker for $\delta(w)\ll1$. In this regime, we write
\begin{equation}\label{q*_w>1}
    q^{*}=\frac{1}{w}-\delta(w),\quad\delta(w)\ll1.
\end{equation}
Substitution into Eq.~\eqref{eq_q} yields
\begin{equation}\label{delta_FJN_w>1}
    \delta(w)\approx\frac{1}{2w^2+3w}.
\end{equation}
The free energy density then reads
\begin{equation}\label{psi_FJN_w>1}
    \psi=k_{\rm{B}}T\ln\left(\frac{2w+2}{2w^2+3w}\right),
\end{equation}
and the mean fraction of bound cross-links is given by
\begin{equation}\label{n_c_FJN_w>1}
    \langle n_{\rm{c}}\rangle=1-\frac{w}{2w^2+5w+3}.
\end{equation}
Its plot as a function of the dimensionless force $bf/k_{\rm{B}}T$ is shown in Fig.~\subref*{fig:n_c_FJN_w>1}. In this case, the plotted range is restricted from below by the condition $w>1$, that is, $bf/k_{\rm{B}}T>N_0/\sigma\exp(\varepsilon/k_{\rm{B}}T)$. The practical cutoff is again chosen heuristically to lie safely above this lower limit, ensuring $\delta(w)\ll1$ throughout [Fig.~\subref*{fig:n_c_FJN_overall}]. Equally important, the range is also restricted from above by the smallness condition $a\ll N_0\,{\ell}^2$, i.e., $bf/k_{\rm{B}}T\ll N_0/2\pi\sigma$. Thus, the force cannot be increased arbitrarily but is limited by the validity condition for the necklace model. Within the strong stretching regime, Eq.~\eqref{n_c_FJN_w>1} describes the mean fraction of bound cross-links in a directed configuration of the FJN associated with a higher level of stretching intensity, capturing the culmination of strong stretching behavior. Reflecting its high binding affinity, we refer to this regime as the strongly bound regime. 

A more detailed exposition of $\langle n_{\rm{c}}\rangle$ in both regimes is deferred to the next subsection, where it is discussed in the broader context of comparison with the wormlike counterpart.

\subsection{Wormlike necklace}\label{WLN}

For the sake of a parallel presentation, we begin by introducing the model and formulating its effective Hamiltonian, though the analysis will ultimately proceed via the Gaussian slinky representation. The wormlike necklace (WLN) consists of a pair of identical wormlike chains, with cross-linking sites equally spaced at positions $kL_0$ ($k=1,2,3,\cdots,\nu$). Each site can undergo reversible binding and unbinding. Under symmetric stretching, the effective Hamiltonian in the weakly bending regime [cf.~Eq.~\eqref{H_cWLL'}] is given by
\begin{align}\label{H_WLN}
    \mathscr{H}_{\rm{WLN}}&=\frac{\kappa}{2}\int_{0}^{\nu L_0}ds\left[\left(\frac{\partial\bm{t}_{1,\perp}}{\partial s}\right)^2+\left(\frac{\partial\bm{t}_{2,\perp}}{\partial s}\right)^2\right]\nonumber\\
    &\quad+\frac{\xi}{2}\sum_{k=1}^{\nu}\theta_k\left(\int_{0}^{kL_{0}}ds\left(\bm{t}_{1,\perp}-\bm{t}_{2,\perp}\right)\right)^2-\sum_{k=1}^{\nu}\theta_k\mu\nonumber\\
    &\quad-2\nu L_0f+\frac{f}{2}\int_{0}^{\nu L_0}ds\left({\bm{t}_{1,\perp}}^2+{\bm{t}_{2,\perp}}^2\right),
\end{align}
where the binary variable $\theta_{k}$ and the energy parameter $\mu$ are defined as in the FJN model. While Eq.~\eqref{H_WLN} provides a complete description, the projected picture grounded in transverse fluctuations offers a more tractable framework for analyzing the mean fraction of bound cross-links. In what follows, we develop the Gaussian slinky representation for the WLN model. To proceed, we assume that the basic contour length $L_0$ is much larger than the memory length ($L_0\gg l_{\rm{m}}$); otherwise the application of two-dimensional Gaussian statistics for the transverse fluctuations would break down. Accordingly, our analysis of the WLN is restricted to the weakly bending regime realized under strong stretching forces rather than by a large persistence length.

We first consider a single WLC under tension in the weakly bending regime. The transverse fluctuations of the weakly bent WLC are given by $\langle{\bm{R}_{\perp}}^2\rangle_{\rm{WLC}}=2k_{\rm{B}}TL/f$. For our purposes, we assume that the chain's contour length is much larger than its memory length ($L\gg l_{\rm{m}}$). In this regime, the WLC can be viewed as a concatenation of rigid segments, each of length $\widetilde{b}=2l_{\rm{m}}$. The number of such segments is then identified as $N=L/\widetilde{b}$. Using $N$ as the size parameter in place of $L$, the transverse fluctuations can be recast as
\begin{equation}\label{fluctuation_t_WLC'}
    \langle{\bm{R}_{\perp}}^2\rangle_{\rm{WLC}}=\frac{2k_{\rm{B}}TN\widetilde{b}}{f}=N\sqrt{\frac{16{k_{\rm{B}}}^3T^3l_{\rm{p}}}{f^3}}.
\end{equation}
Thus, similar to the FJC case, the transverse trajectory of the WLC follows the statistics of a two-dimensional random walk and becomes Gaussian for $N\gg 1$. The apparent difference lies in their effective step sizes (or effective bond lengths in their Gaussian chain representations):
\begin{align}\label{ell'_1}
    \ell&=\left(\frac{2k_{\rm{B}}Tb}{f}\right)^{1/2}\quad\rm{(FJC)},\nonumber\\
    \ell^{\prime}&=\left(\frac{16{k_{\rm{B}}}^3T^3l_{\rm{p}}}{f^3}\right)^{1/4}\quad\rm{(WLC)}.
\end{align}
Note that $\ell^{\prime}$ also corresponds to the root-mean-square transverse excursion of an individual segment in the equivalent freely jointed picture. Denoting the local tilt angle by $\theta$, the projected length of $\widetilde{b}$ is
\begin{equation}\label{ell'_2}
    \sqrt{\left\langle\left(\widetilde{b}\sin\theta\right)^2\right\rangle}\approx\sqrt{\left(2l_{\rm{m}}\right)^2\langle{\bm{t}_{\perp}}^2\rangle}=\left(\frac{16{k_{\rm{B}}}^3T^3l_{\rm{p}}}{f^3}\right)^{1/4}.
\end{equation}

Once the Gaussian slinky model has been established for the FJN, its extension to the WLN requires only a straightforward modification of the statistical weight that determines the observables of interest. Here, it is important to note that $N_0=L_0/\widetilde{b}$ is force-dependent, unlike in the FJN case. To isolate the effects of force dependence, the statistical weight must be expressed in terms of $L_0$, which is the fixed spacing between cross-linking sites in the construction of the WLN model. In the FJN model, the spacing between cross-linking sites is parameterized by a constant $N_0$ from the outset, which is equivalent to fixing $L_0=N_0b$. In the Gaussian slinky representation of the WLN, the partition function of the loop of size $kL_0$ is obtained by substituting $\ell^{\prime}$ for $\ell$ in Eq.~\eqref{Z_GL}, and then restoring $L_0$. For a side-by-side comparison, we present the partition functions for both models:
\begin{align}\label{Z_GL'}
    \mathscr{Z}(kN_0)&=\frac{1}{2\pi}\left(\frac{a}{{\ell}^2}\right)(kN_0)^{-1}\exp\left(\frac{\varepsilon}{k_{\rm{B}}T}\right)\nonumber\\
    &=\frac{af}{4\pi k_{\rm{B}}T}(kN_0b)^{-1}\exp\left(\frac{\varepsilon}{k_{\rm{B}}T}\right)\quad\rm{(FJN)},\nonumber\\
    \mathscr{Z}^{\prime}(kL_0)&=\frac{1}{2\pi}\left(\frac{a}{{\ell^{\prime}}^2}\right)(kN_0)^{-1}\exp\left(\frac{\varepsilon}{k_{\rm{B}}T}\right)\nonumber\\
    &=\frac{af}{4\pi k_{\rm{B}}T}(kL_0)^{-1}\exp\left(\frac{\varepsilon}{k_{\rm{B}}T}\right)\quad\rm{(WLN)}.
\end{align}
The decisive statistical weight is given by the partition function of the elementary loop (of size $N_0$ for the FJN and $L_0$ for the WLN):
\begin{align}\label{w'}
    w&=\frac{af}{4\pi k_{\rm{B}}TN_0b}\exp\left(\frac{\varepsilon}{k_{\rm{B}}T}\right)\quad\rm{(FJN)},\nonumber\\
    w^{\prime}&=\frac{af}{4\pi k_{\rm{B}}TL_0}\exp\left(\frac{\varepsilon}{k_{\rm{B}}T}\right)\quad\rm{(WLN)}.
\end{align}
Interestingly, the partition function of the loop (and thus the elementary statistical weight) turns out to be identical in both the freely jointed and wormlike models. This may seem remarkable at first glance, given that the effective step size of the projected trajectory is model-specific, scaling differently with force: $\ell\sim f^{-1/2}$ versus $\ell^{\prime}\sim f^{-3/4}$ [see Eq.~\eqref{ell'_1}]. In fact, the partition function in the projected picture fundamentally stems from the transverse fluctuations of the single chain, which happen to be identical across the models: $2k_{\rm{B}}TNb/f$ for the FJC and $2k_{\rm{B}}TL/f$ for the WLC. These fluctuations enter the two-dimensional Gaussian distribution in Eq.~\eqref{rho} in the collective form of $N\ell^2$ (or $N{\ell^{\prime}}^2$), which is the only ingredient that differs between the two models. As a result, the same transverse fluctuations produce the same distribution, leading to the same looping probability [Eq.~\eqref{P_l}] and, consequently, the same partition function.

\begin{figure*}
    \sidesubfloat[]{\includegraphics[width=0.28\textwidth]{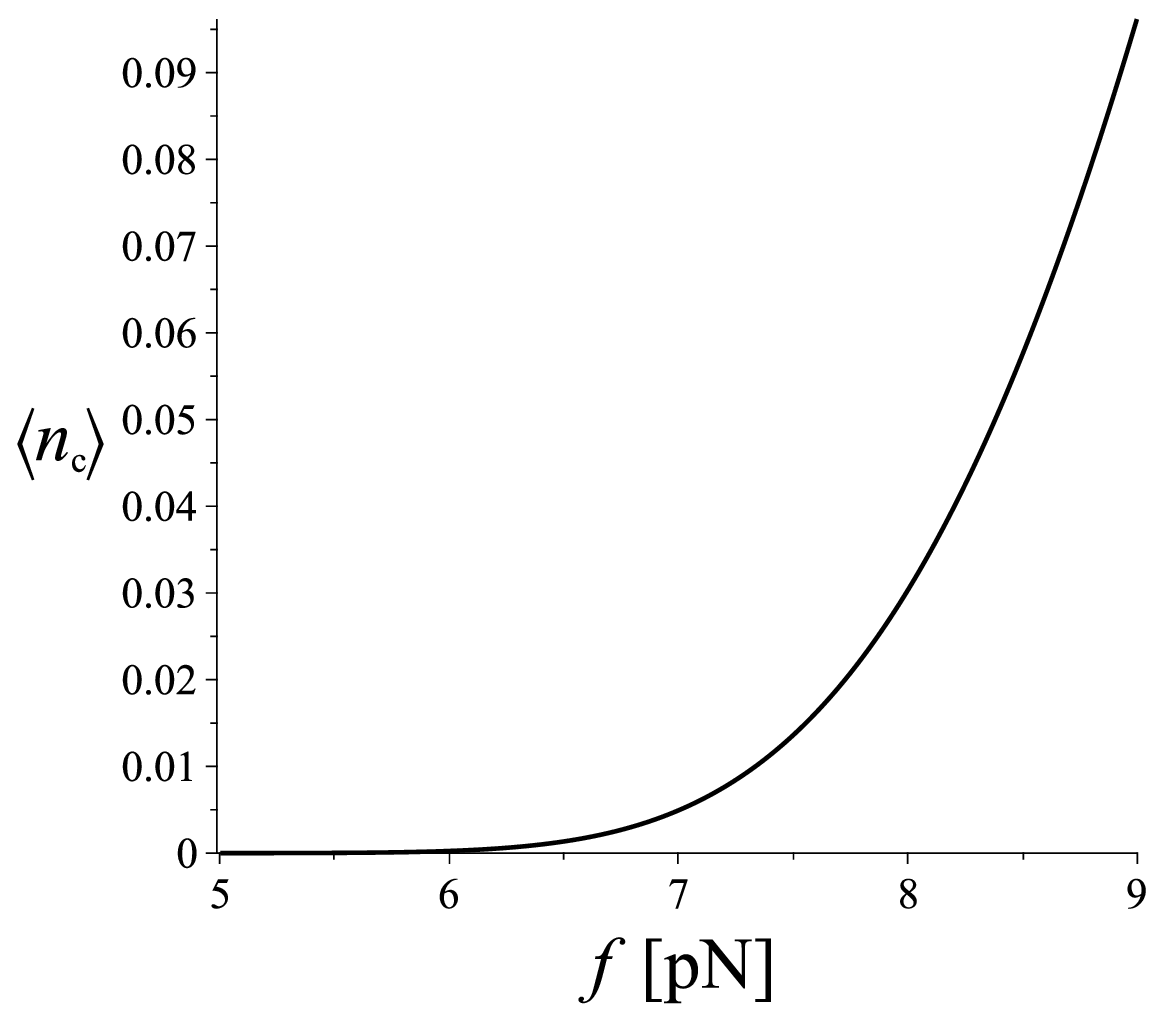}\label{fig:n_c_WLN_w<1}}
    \hfill
    \sidesubfloat[]{\includegraphics[width=0.28\textwidth]{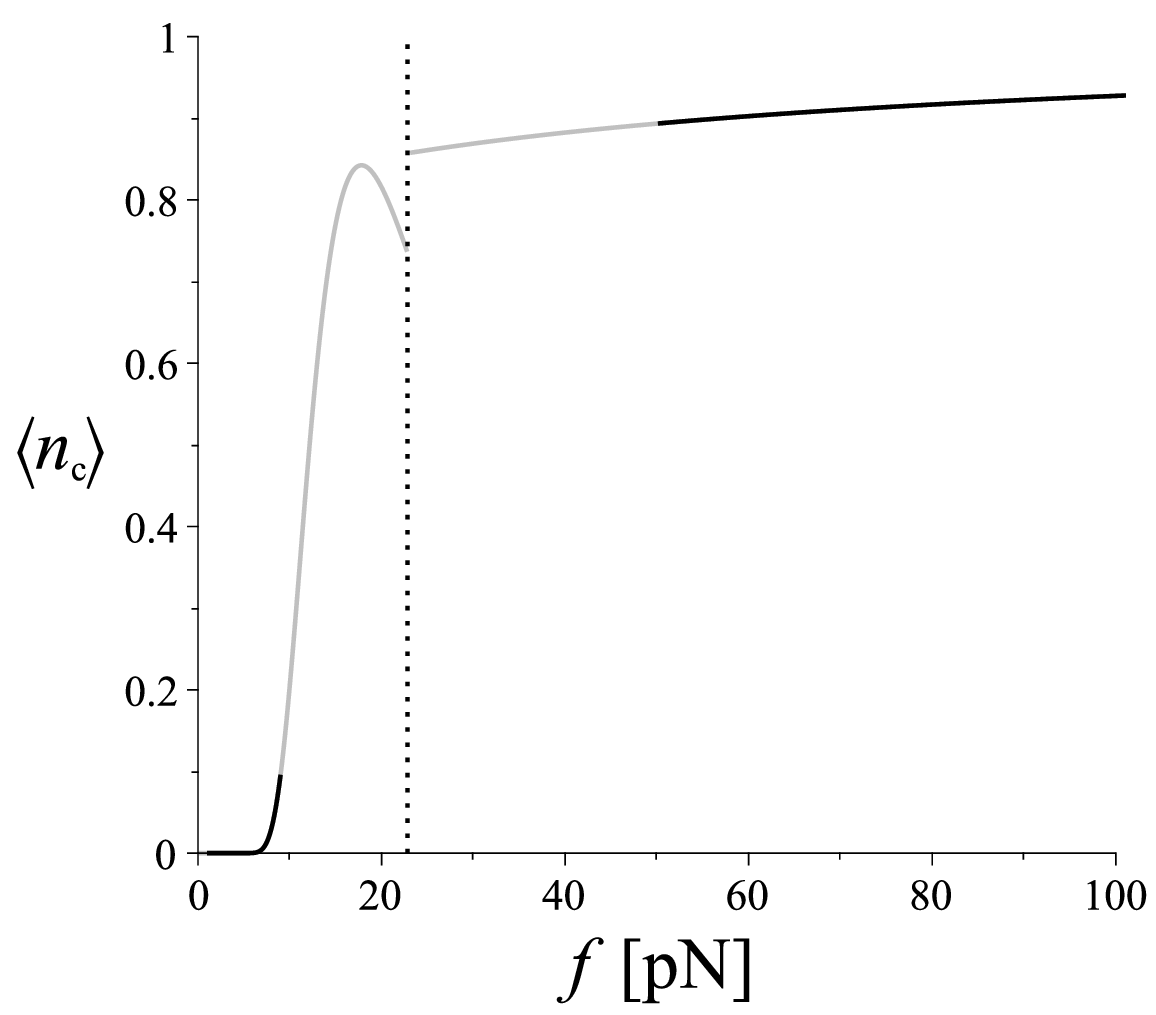}\label{fig:n_c_WLN_overall}}
    \hfill
    \sidesubfloat[]{\includegraphics[width=0.28\textwidth]{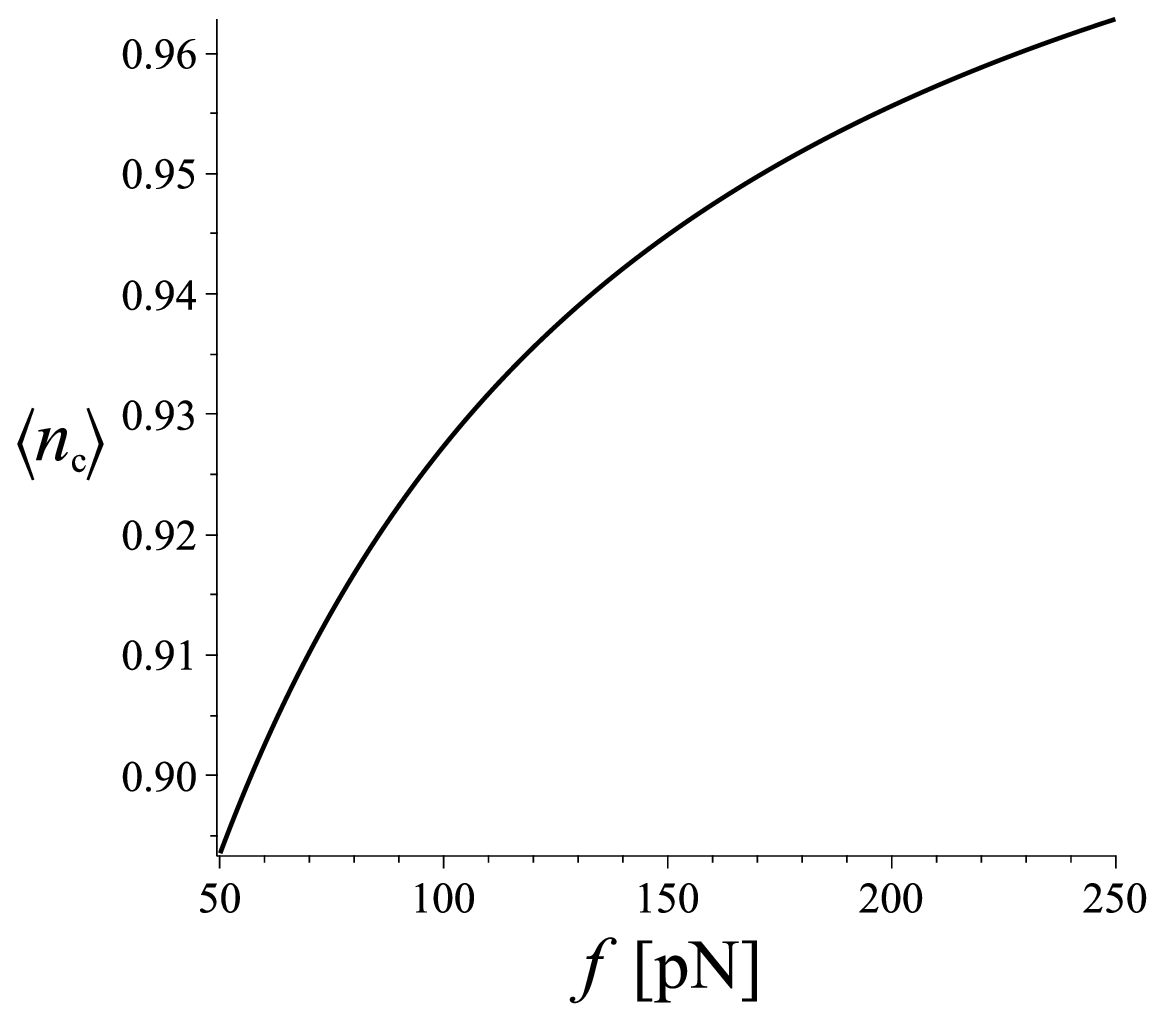}\label{fig:n_c_WLN_w>1}}
    \caption{Mean fraction of bound cross-links $\langle n_{\rm{c}}\rangle$ as a function of the actual force $f$, shown for parameter values $L_0=\SI{1}{\micro\metre}$, $a=\SI{0.1}{\nano\metre\squared}$, $T=\SI{300}{\kelvin}$ ($k_{\rm{B}}T\approx\SI{4}{\pico\newton\nano\metre}$), and $\varepsilon/k_{\rm{B}}T=10$. The results are model-independent: WLN and FJN yield identical curves for the given parameter values.  (a) Weakly bound regime ($w^{\prime}<1$). (b) The vertical dotted line marks the reference point $w^{\prime}=1$, corresponding to $f\approx\SI{23}{\pico\newton}$. The gray segments, which exhibit an unphysical peak, lie outside the domain of validity of the asymptotic approximation and are shown only to illustrate the overall trend. The reliable region (black segments) is determined heuristically and should be regarded as the physically meaningful prediction. (c) Strongly bound regime ($w>1$). The model's validity condition, $f\ll2k_{\rm{B}}TL_0/a=\SI{8e4}{\pico\newton}$, prevents arbitrarily large forces.}
    \label{fig:n_c_WLN}
\end{figure*}

We now dive directly into the results for the WLN without further preamble. The mean fraction of bound cross-links in each asymptotic regime has the same form as in the FJN case [Eqs.~\eqref{n_c_FJN_w<1} and~\eqref{n_c_FJN_w>1}], with the substitution $w\to w^{\prime}$. In the weakly bound regime ($w^{\prime}<1$), it is given by
\begin{equation}\label{n_c_WLN_w'<1}
    \langle n_{\rm{c}}\rangle=\frac{\left(\frac{2}{{w^{\prime}}^2}-\frac{1}{{w^{\prime}}}\right)\exp\left(-\frac{1}{{w^{\prime}}^2}+\frac{1}{{w^{\prime}}}-1\right)}{1-\exp\left(-\frac{1}{{w^{\prime}}^2}+\frac{1}{{w^{\prime}}}-1\right)},
\end{equation}
and in the strongly bound regime ($w^{\prime}>1$), it becomes 
\begin{equation}\label{n_c_WLN_w'>1}
    \langle n_{\rm{c}}\rangle=1-\frac{{w^{\prime}}}{2{w^{\prime}}^2+5{w^{\prime}}+3}.
\end{equation}
As functions of $f$, these expressions can be approximated as
\begin{equation}\label{n_c_WLN_f}
    \langle n_{\rm{c}}\rangle\approx
        \begin{cases}
            \displaystyle
            \frac{2{k_{\rm{B}}}^2T^2}{c^2f^2}\exp\left(-\frac{{k_{\rm{B}}}^2T^2}{c^2f^2}+\frac{k_{\rm{B}}T}{c f}-1\right), & w^{\prime}<1,\\[3ex]
            \displaystyle
            1-\frac{k_{\rm{B}}T}{2c f+5k_{\rm{B}}T}, & w^{\prime}>1,
        \end{cases}
\end{equation}
where $c\equiv a\exp(\varepsilon/k_{\rm{B}}T)/4\pi L_0$ has dimension of length. Their plots are shown in Figs.~\subref*{fig:n_c_WLN_w<1} and~\subref*{fig:n_c_WLN_w>1}, respectively. This time, we plot the $\langle n_{\rm{c}}\rangle$ curves against the actual force $f$ in piconewtons. For example, when the polymer's stiffness is characterized by $l_{\rm{p}}=\SI{50}{\nano\metre}$, it has $\langle X\rangle_{\rm{WLC}}/L\approx90\%$ at $f=\SI{2}{\pico\newton}$ and $T=\SI{300}{\kelvin}$, illustrating the relevant force scale. As seen in the FJN curves, the mean occupation profiles display a partially sigmoidal behavior, with two segmented portions---one concave and the other convex. In the weakly bound regime [Fig.~\subref*{fig:n_c_WLN_w<1}], the mean occupation number increases monotonically with force, as stronger stretching promotes alignment---and consequently binding---by suppressing transverse fluctuations. It also exhibits a force-induced crossover from $\langle n_{\rm{c}}\rangle\approx0$ (nearly unbound necklace) to appreciable values (partially bound necklace), indicative of cooperative cross-link binding. In the strongly bound regime [Fig.~\subref*{fig:n_c_WLN_w>1}], the mean occupation number asymptotically approaches unity (fully bound necklace) as the force increases. However, the force cannot be increased arbitrarily, as the applicability of the necklace model requires $f\ll2k_{\rm{B}}TL_0/a$. Because most cross-links are already bound at an earlier stage and thus the boost from binding cooperativity is largely exhausted, additional force produces only minor increases, resulting in gradual saturation.

For the FJN with the same basic spacing $L_0$ as the WLN, the mean occupation number is likewise described by Eqs.~\eqref{n_c_WLN_w'<1}--\eqref{n_c_WLN_f}. By contrast, the force--extension relations of the cross-linked loops (or single chains) differ qualitatively between the two models. The freely jointed system is more extended than the wormlike system at the same force and approaches full extension more readily as force increases, with $\langle X\rangle_{\rm{cFJL}}/L\sim1-\mathcal{O}(f^{-1})$ versus $\langle X\rangle_{\rm{cWLL}}/L\sim1-\mathcal{O}(f^{-1/2})$. In a naive geometric picture, this greater longitudinal alignment in the freely jointed case would imply a smaller transverse separation between the chain pair, which in turn seems favorable for binding in the necklace configuration. This apparent conflict between the given result and the naive expectation can be resolved by directly calculating the average amplitude of the transverse fluctuations along the chain's contour. Since the transverse fluctuations of the pair are simply twice those of an individual chain, we consider the simpler case of a single chain. For the FJC, the statistical independence of transverse fluctuations in each segment, $\langle(\bm{R}_{\perp,n})^2\rangle=2nk_{\rm{B}}Tb/f$, leads to the contour-averaged transverse amplitude
\begin{equation}\label{fluctuation_t_FJC_avg}
    \frac{1}{N}\sum_{n=1}^{N}\langle(\bm{R}_{\perp,n})^2\rangle=\frac{(N+1)k_{\rm{B}}Tb}{f},
\end{equation}
which, for $N\gg1$, is well approximated as $Nk_{\rm{B}}Tb/f=k_{\rm{B}}TL/f$. For the WLC, the contour-averaged transverse amplitude is given by [see Eqs.~\eqref{app:fluctuation_t_WLC_avg_1}--\eqref{app:fluctuation_t_WLC_avg_3}]
\begin{align}\label{fluctuation_t_WLC_avg}
    \frac{1}{L}\int_{0}^{L}ds\left\langle{\left[{\bm{R}_{\perp}}(s)\right]}^2\right\rangle&=\frac{k_{\rm{B}}TL}{f}-\frac{k_{\rm{B}}Tl_{\rm{m}}}{f}\coth\left(\frac{L}{l_{\rm{m}}}\right)\nonumber\\
    &\quad+\frac{k_{\rm{B}}T{l_{\rm{m}}}^2}{Lf},
\end{align}
which, for $L\gg l_{\rm{m}}$, reduces to
\begin{align}\label{fluctuation_t_WLC_avg_approx}
    \frac{1}{L}\int_{0}^{L}ds\left\langle{\left[{\bm{R}_{\perp}}(s)\right]}^2\right\rangle&\approx\frac{k_{\rm{B}}TL}{f}\left[1-\frac{l_{\rm{m}}}{L}+\left(\frac{l_{\rm{m}}}{L}\right)^2\right]\nonumber\\
    &\approx\frac{k_{\rm{B}}TL}{f}.
\end{align}
It is now clear that the reasoning based solely on force--extension relations is misleading. Despite the difference in their average longitudinal extensions, the freely jointed and wormlike models exhibit the same average transverse fluctuations. In both models, the contour-averaged amplitude of the transverse fluctuations is half the endpoint amplitude, as measured by $\langle{\bm{R}_{\perp}}^2\rangle=2k_{\rm{B}}TL/f$. Since these transverse fluctuations govern the binding statistics, their equality in the two models necessarily yields the same mean occupation number.

\subsection{Crossover estimation}\label{sec:crossover}

The partially sigmoidal shape of the mean occupation curves indicates a crossover between the weakly and strongly bound regimes. While direct information about the intermediate regime is not available, the crossover can be inferred by matching the free energy densities of the two asymptotic regimes. As the binding behavior is identical in the FJN and WLN models, we treat both collectively as the necklace, using $L_0$ and expressing $w=w(L_0)$ rather than $w=w(N_0b)$ to place them on equal footing. 

From Eqs.~\eqref{psi_FJN_w<1} and~\eqref{psi_FJN_w>1}, the crossover is estimated to occur over a range of forces around $f\approx f_{\rm{c}}$, with the corresponding value $w_{\rm{c}}$ determined approximately by
\begin{equation}\label{crossover}
    1-\exp\left(-\frac{1}{{w_{\rm{c}}}^2}+\frac{1}{w_{\rm{c}}}-1\right)\approx\frac{2w_{\rm{c}}+2}{2{w_{\rm{c}}}^2+3w_{\rm{c}}}.
\end{equation}
The numerical solution $w_{\rm{c}}\approx1.6$ (in comparison with the heuristic reference point $w=1$) gives
\begin{equation}\label{f_c_necklace}
    f_{\rm{c}}\approx\frac{6.4\pi k_{\rm{B}}TL_0}{a\exp(\varepsilon/k_{\rm{B}})}\sim\frac{\pi k_{\rm{B}}TL_0}{a\exp(\varepsilon/k_{\rm{B}})},
\end{equation}
which sets the characteristic force scale around which the necklace crosses over from the weakly to the strongly bound regime.

We note that the crossover depends strongly on the value of $\varepsilon$, which in turn is determined by the depth $\varepsilon_a$ of the confining potential well [see Eq.~\eqref{epsilon}]. If the well is deep compared with the thermal energy $k_{\rm{B}}T$, the argument $\varepsilon/k_{\rm{B}}T$ of the  exponential factor in Eq.~\eqref{w'} can become large, such that only the strongly bound regime (i.e., $w>1$ for all $f$) is realized and $\langle n_{\rm{c}}\rangle\approx1$ under strong tension within the model's validity condition, $f\ll2k_{\rm{B}}TL_0/a$. In contrast, if the well is shallow compared with $k_{\rm{B}}T$, the exponent $\varepsilon/k_{\rm{B}}T$ can be small or may even become negative, such that only the weakly bound regime (i.e., $w<1$ for all $f$) is realized and $\langle n_{\rm{c}}\rangle\approx0$ under strong tension within the same validity range. This observation naturally prompts the question: do the same tendencies persist into the large-force regime beyond the model's validity range? In principle, the force can grow arbitrarily large; the restriction $f\ll2k_{\rm{B}}TL_0/a$ arises solely from the construction of the necklace model. In the deep-well case, as well as in the intermediate-depth case (partially sigmoidal), the answer is expected to be affirmative, since there is no physical reason for $\langle n_{\rm{c}}\rangle$ to destabilize or suddenly decrease. However, in the shallow-well case it remains uncertain whether the same tendency (weak binding) continues, and it is in fact expected not to persist indefinitely. This motivates a complementary analysis free from the force restriction, allowing us to explore the \textit{terra incognita} of the shallow-well case.

\subsection{Continuum limit and quantum analogy}\label{sec:quantum}

\begin{figure}
    \includegraphics[width=8.6cm]{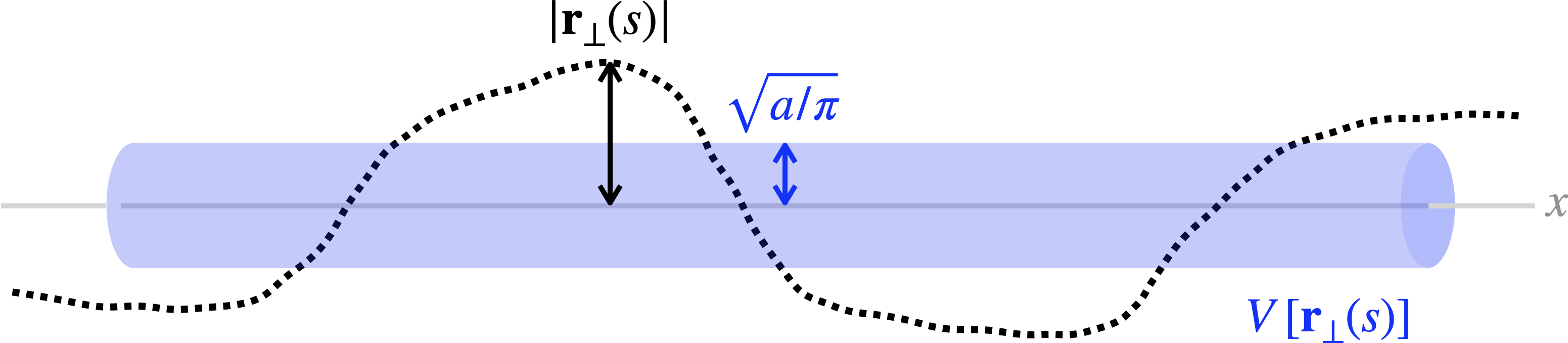}
    \caption{Schematic illustration of the continuum limit, where the relative transverse trajectory $\bm{r}_{\perp}(s)$ of the necklace is confined by the effective potential well $V[\bm{r}_{\perp}(s)]$.}
    \label{fig:CL}
\end{figure}

An alternative to the two-state sequence picture of the polymer necklace is to exploit the well-known mapping between polymer trajectories and the imaginary-time paths of a fictitious quantum particle, a formal correspondence well established in polymer theory \cite{Edwards,deGennes3,Nelson1,PB2}. Guided by this quantum mechanical analogy, and following in particular Refs.~\cite{Nelson1,PB2} on directed polymers, we address the shallow-well case hitherto unresolved. Before embarking on the quantum analogy, we note that $\langle n_{\rm{c}}\rangle$ depends neither on the chain model (freely jointed or wormlike) nor on bending stiffness. Thus, we treat the chains as flexible directed polymers \cite{PB2} by neglecting the bending potential ($\kappa=0$) in Eq.~\eqref{H_WLN}. This treatment is coherent with the WLN assumption that the weakly bending (directed) regime is realized under strong stretching, where stretching energy dominates over bending, rather than by a large persistence length. This modeling choice also makes the subsequent analogy well suited to our context and transparent.

As a first step, we perform a change of variables in the two-chain description of the necklace by introducing the ``center-of-mass'' (midpoint) transverse trajectory 
\begin{equation*}
    \bm{R}_{\perp}(s)\equiv\frac{\bm{r}_{1,\perp}(s)+\bm{r}_{2,\perp}(s)}{2},
\end{equation*}
and the relative transverse trajectory
\begin{equation*}
    \bm{r}_{\perp}(s)\equiv\bm{r}_{1,\perp}(s)-\bm{r}_{2,\perp}(s),
\end{equation*}
with $s\in[0,L]$. This change will later enable the partition function to decouple into contributions from $\bm{R}_{\perp}(s)$ and $\bm{r}_{\perp}(s)$, thereby reducing the two-chain picture to a single-chain description and rendering the quantum analogy straightforward, as will be shown shortly. In terms of these variables, the stretching energy of the directed necklace becomes
\begin{equation}\label{T}
\mathscr{T}=-LF+\frac{F}{2}\int_{0}^{L}ds\left(\frac{\partial\bm{R}_{\perp}}{\partial s}\right)^2+\frac{\widetilde{f}}{2}\int_{0}^{L}ds\left(\frac{\partial\bm{r}_{\perp}}{\partial s}\right)^2,
\end{equation}
where $F\equiv2f$ and $\widetilde{f}\equiv f/2$. Here, the force plays the role of mass in a two-body system: $F$ corresponds to the total mass, while $\widetilde{f}$ corresponds to the reduced mass.

For the reversible two-state cross-linking interaction, we first focus on a single site at $s=kL_0$. In the adopted coordinates, the cross-linking potential energy depends only on the relative displacement $\bm{r}_{\perp}(kL_0)$:
\begin{equation}\label{U}
    \mathscr{U}_k=
        \begin{cases}
            \frac{\xi}{2}\left[\bm{r}_{\perp}(kL_0)\right]^2-\mu, & \theta_k=1,\\[1ex]
            0, & \theta_k=0.
        \end{cases}
\end{equation}
Constructing the partition function for a two-state cross-linking site then yields the effective potential
\begin{equation}
    \mathscr{U}_{\rm{eff},k}=-k_{\rm{B}}T\,\ln\!\left[1+\exp\left(-\frac{\xi\left[\bm{r}_{\perp}(kL_0)\right]^2}{2k_{\rm{B}}T}+\frac{\mu}{k_{\rm{B}}T}\right)\right].
\end{equation}
In practice, this interaction may be viewed as an effective short-range confining potential acting on the relative displacement $\bm{r}_{\perp}(kL_0)$. For simplicity, we model it as a circularly symmetric potential well of the form
\begin{equation}\label{U_eff}
    \mathscr{U}_{\rm{eff},k}=-U_0\,\theta\left(\Delta-\lvert\bm{r}_{\perp}(kL_0)\rvert\right),
\end{equation}
where $U_0$ is the depth, $\Delta$ the range, and $\theta(x)$ the Heaviside step function. The total interaction energy is obtained by summing over all cross-linking sites, $\sum_{k}\mathscr{U}_{\rm{eff},k}$. Because the present description, like the Gaussian slinky model, involves confinement in the transverse plane, we retain the same physics across both approaches by identifying the potential parameters as $U_0=\varepsilon_a$ and $\Delta=\sqrt{a/\pi}$. It is important to note, however, that unlike in the Gaussian slinky model, this assignment entails no restriction on the stretching force. Thus, we can benefit from the consistency of parameters without inheriting the force limitation---a key step toward exploring the shallow-well case at large forces, previously inaccessible within the Gaussian slinky representation. 

For a large number of equally spaced cross-linking sites, the discrete array of potential wells may be approximated by a continuous potential channel along the force axis. In this continuum limit ($\sum_{k}\to\int_{0}^{L}ds/L_0$), the interaction is described by an effective potential density (an effective potential per unit contour length) that confines the relative transverse trajectory (see Fig.~\ref{fig:CL}):
\begin{equation}\label{V}
    V[\bm{r}_{\perp}(s)]=-\frac{\varepsilon_a}{L_0}\,\theta\left(\sqrt{\frac{a}{\pi}}-\lvert\bm{r}_{\perp}(s)\rvert\right).
\end{equation}
It should be noted, however, that this continuum-limit description is not appropriate for the deep-well case. This mean-field-type treatment amounts to smearing the discrete cross-linking events into a coarse-grained, smooth, linear potential density. When the potential wells are shallow compared with the thermal energy $k_{\rm{B}}T$, this averaging is reasonable and still captures the essential physics. By contrast, when the wells are deep compared with $k_{\rm{B}}T$, cross-linking becomes strongly localized, and thus the discrete nature of the interaction is crucial. In other words, strong binding requires treating the potential as a set of localized wells rather than as a smooth, continuous channel. Moreover, the confining potential originates from the chain interactions themselves, not from an independent external field. As such, the interaction is intrinsically coupled to the chain trajectories. In the shallow-well case, the trajectories are influenced collectively by the array of cross-linking sites, with the overall behavior dominated by the external stretching force rather than by individual binding events. By contrast, in the deep-well case, the physics is site-specific rather than collective: the chain experiences strong localized trapping (or pinning) at periodically spaced individual sites. In this case, chain trajectories cannot be treated as if they were subjected to a smooth, continuous background field. This distinction explains why the mean-field-like treatment inherent in the continuum limit is adequate for the shallow-well case but inadequate for the deep-well case.

\begin{widetext}
    The partition function of the system under consideration is given by the path integral
    \begin{equation}\label{Z}
        \mathscr{Z}=\int\mathscr{D}[\bm{R}_{\perp}(s)]\int\mathscr{D}[\bm{r}_{\perp}(s)]\exp\left\{\frac{LF}{k_{\rm{B}}T}-\frac{1}{k_{\rm{B}}T}\int_{0}^{L}ds\left[\frac{F}{2}\left(\frac{\partial\bm{R}_{\perp}(s)}{\partial s}\right)^2+\frac{\widetilde{f}}{2}\left(\frac{\partial\bm{r}_{\perp}(s)}{\partial s}\right)^2+V[\bm{r}_{\perp}(s)]\right]\right\},
    \end{equation}
    where $\int\mathscr{D}[\bm{R}_{\perp}(s)]$ and $\int\mathscr{D}[\bm{r}_{\perp}(s)]$ denote the sums over all possible trajectories connecting $\bm{R}_{\perp}(0)$ to $\bm{R}_{\perp}(L)$ and $\bm{r}_{\perp}(0)$ to $\bm{r}_{\perp}(L)$, respectively. To factor out the entropic contribution from unconfined (non-interacting) trajectories, we consider the relative partition function given by the ratio of the full partition function to that of the free system:
    \begin{equation}\label{Z_relative}
        \mathscr{Z}_{\rm{rel}}=\frac{\int\mathscr{D}[\bm{r}_{\perp}(s)]\exp\left\{-\frac{1}{k_{\rm{B}}T}\int_{0}^{L}ds\left[\frac{\widetilde{f}}{2}{\dot{\bm{r}}_{\perp}}^2(s)+V[\bm{r}_{\perp}(s)]\right]\right\}}{\int\mathscr{D}[\bm{r}_{\perp}(s)]\exp\left\{-\frac{1}{k_{\rm{B}}T}\int_{0}^{L}ds\left[\frac{\widetilde{f}}{2}{\dot{\bm{r}}_{\perp}}^2(s)\right]\right\}},
    \end{equation}
    where the path-independent factor $\exp(LF/k_{\rm{B}}T)$ and the separable center-of-mass path integral cancel out, and we define ${\dot{\bm{r}}_{\perp}}^2(s)\equiv\partial\bm{r}_{\perp}(s)/\partial s$. Eventually, only the relative motion matters, and the two-chain necklace system reduces to an effective single-chain picture in $(1+2)$ dimensions.
\end{widetext}

Equation~\eqref{Z_relative} can be viewed as the Feynman path integral representation of an (unnormalized) density matrix element for a quantum particle in the position basis (obtained via Wick rotation, $t\to-i\tau$) \cite{Feynman1,Feynman2,Zee}. The parameters of the directed polymer system map onto those of a quantum system as $k_{\rm{B}}T\longleftrightarrow\hbar$, $\widetilde{f}\longleftrightarrow m$, $s\longleftrightarrow\tau$, and $L\longleftrightarrow\beta\hbar$, where $m$ is the mass, $\tau$ the imaginary time, and $\beta\hbar$ the total imaginary-time interval, with $\beta$ the inverse temperature of the quantum system in energy units. In this analogy, the polymer trajectory corresponds to the motion of a fictitious quantum particle in imaginary time. Thus, in the language of Schrödinger wave mechanics, Eq.~\eqref{Z_relative} can be expressed as the ratio of the associated propagators,
\begin{equation}
    \mathscr{Z}_{\rm{rel}}=\frac{\sum_{n}\phi_n(\bm{r}_{\perp};L)\,\phi_n^*(\bm{r}_{\perp};0)\exp\left(\frac{-\lambda_nL}{k_{\rm{B}}T}\right)}{\left(\frac{\widetilde{f}}{2\pi k_{\rm{B}}TL}\right)\exp\left(-\frac{\widetilde{f}\left[\bm{r}_{\perp}(L)-\bm{r}_{\perp}(0)\right]^2}{2k_{\rm{B}}TL}\right)},
\end{equation}
where the eigenfunctions $\phi_n(\bm{r}_{\perp})$ with eigenvalues $\lambda_n$ satisfy the contour-length independent Schrödinger-like equation
\begin{equation}\label{Schrödinger}
    \left(-\frac{(k_{\rm{B}}T)^2}{2\widetilde{f}}\nabla_{\perp}^2+V(\bm{r}_{\perp})\right)\phi_n(\bm{r}_{\perp})=\lambda_n\phi_n(\bm{r}_{\perp}),
\end{equation}
and the denominator results from the propagator for a two-dimensional free particle. As $L\to\infty$, corresponding to $\beta\to\infty$ in the quantum analogue, the numerator is dominated by the term with the lowest eigenvalue $\lambda_0$, since contributions from higher eigenvalues are exponentially suppressed:
\begin{align}\label{lambda}
    &\sum_{n}\phi_n(\bm{r}_{\perp};L)\,\phi_n^*(\bm{r}_{\perp};0)\exp\left(\frac{-\lambda_nL}{k_{\rm{B}}T}\right)\nonumber\\
    &\quad\approx\phi_0(\bm{r}_{\perp};L)\,\phi_0^*(\bm{r}_{\perp};0)\exp\left(\frac{-\lambda_0L}{k_{\rm{B}}T}\right).
\end{align}
Consequently, in the thermodynamic limit, the binding free energy density (binding free energy per unit contour length) is determined by the lowest eigenvalue $\lambda_0$:
\begin{equation}\label{g}
    g=\lim_{L\to\infty}\frac{-k_{\rm{B}}T\ln\mathscr{Z}_{\rm{rel}}}{L}=\lambda_0.
\end{equation}

The behavior of the directed polymer necklace now maps onto the well-known quantum problem of a particle confined in a two-dimensional circular potential well \cite{Galitski,Landau}. In polar coordinates, Eq.~\eqref{Schrödinger} separates into radial and angular parts. The angular dependence is described by plane waves $\exp(im\theta)$ with angular momentum quantum number $m$. Since we are interested in the ground state, which corresponds to zero angular momentum ($m=0$), the problem reduces to the radial equation:
\begin{equation}\label{radial_inside}
    \frac{d^2\phi_0(r_{\perp})}{d\,{r_{\perp}}^2}+\frac{1}{r_{\perp}}\frac{d\,\phi_0(r_{\perp})}{d\,r_{\perp}}+k^2\phi_0(r_{\perp})=0,\quad r_{\perp}<\sqrt{a/\pi},
\end{equation}
\begin{equation}\label{radial_outside}
    \frac{d^2\phi_0(r_{\perp})}{d\,{r_{\perp}}^2}+\frac{1}{r_{\perp}}\frac{d\,\phi_0(r_{\perp})}{d\,r_{\perp}}-{\kappa}^2\phi_0(r_{\perp})=0,\quad r_{\perp}>\sqrt{a/\pi},
\end{equation}
where 
\begin{equation*}
    k^2\equiv\frac{2\widetilde{f}\left[(\varepsilon_a/L_0)-\lvert\lambda_0\rvert\right]}{(k_{\rm{B}}T)^2},\quad\kappa^2\equiv\frac{2\widetilde{f}\lvert\lambda_0\rvert}{(k_{\rm{B}}T)^2}.
\end{equation*}
From the requirements of regularity at the origin and vanishing at infinity, the solutions of these Bessel equations are given by
\begin{equation}\label{Bessel}
    \phi_0(r_{\perp})=
        \begin{cases}
            c_{\rm{in}}\,J_0(kr_{\perp}), & r_{\perp}<\sqrt{a/\pi},\\
            c_{\rm{out}}\,K_0(\kappa r_{\perp}), & r_{\perp}>\sqrt{a/\pi},
        \end{cases}
\end{equation}
where $J_0(x)$ and $K_0(x)$ denote the Bessel and modified Bessel functions of order zero, respectively. The eigenvalue is determined by imposing the continuity of $\phi_0(r_{\perp})$ and its derivative at the boundary $r_{\perp}=\sqrt{a/\pi}$, which gives the transcendental equation
\begin{equation}\label{eigenvalue}
    \frac{kJ_1(k\sqrt{a/\pi})}{J_0(k\sqrt{a/\pi})}=\frac{\kappa K_1(\kappa\sqrt{a/\pi})}{K_0(\kappa\sqrt{a/\pi})}.
\end{equation}

In the weakly bound regime of the shallow-well case, previously observed for $f\ll2k_{\rm{B}}TL_0/a$, the relative displacement $\bm{r}_{\perp}$ is only weakly localized within the potential well $V(\bm{r}_{\perp})$, and the magnitude of the binding free energy density is much smaller than the well depth, $\lvert\lambda_0\rvert\ll\varepsilon_a/L_0$. In this regime, the arguments of the Bessel functions become small, and Eq.~\eqref{eigenvalue} simplifies to
\begin{equation}\label{eigenvalue_w}
    \frac{\varepsilon_a a\widetilde{f}}{2\pi(k_{\rm{B}}T)^2L_0}\ln\left(\frac{\pi(k_{\rm{B}}T)^2}{2a\widetilde{f}\lvert\lambda_0\rvert}\right)\approx1,
\end{equation}
where we have used the asymptotic forms of the Bessel functions for $x\ll1$,
\begin{equation*}
    J_0(x)\approx1,\ J_1(x)\approx\frac{x}{2},\ K_0(x)\approx\ln\left(\frac{1}{x}\right),\ K_1(x)\approx\frac{1}{x}.
\end{equation*}
Thus, the binding free energy density in the weakly bound regime is asymptotically given by
\begin{equation}\label{g_w}
    g_{\rm{w}}\approx-\frac{\pi{k_{\rm{B}}}^2T^2}{af}\exp\left(-\frac{4\pi{k_{\rm{B}}}^2T^2L_0}{\varepsilon_a af}\right),
\end{equation}
and is exponentially small compared with the well depth. Note that a bound (localized) state exists even for a vanishingly shallow potential well. The existence of such a bound state, together with the exponential dependence, is in fact a general feature of one- and two-dimensional potential well problems in quantum mechanics \cite{Landau}.

The crossover behavior in the shallow-well case is already manifest in the form of the binding free energy density. The exponential dependence in Eq.~\eqref{g_w} defines a characteristic force scale
\begin{equation}\label{f_characteristic}
    f^{*}=\frac{\pi{k_{\rm{B}}}^2T^2L_0}{\varepsilon_a a},
\end{equation}
which governs the exponential sensitivity of $g_{\rm{w}}$ to the applied force $f$. Here we point out that, within this quantum analogy, $f$ can become arbitrarily large. For $f\ll f^*$, the exponential factor becomes vanishingly small, leading to extremely weak binding (localization) that corresponds to the shallow-well asymptotic regime. When $f$ reaches values comparable to $f^{*}$ ($f\gtrsim f^{*}$), the exponential factor becomes appreciable, signaling the breakdown of the weakly bound regime and the onset of strong binding (localization). Thus, $f^*$ provides a natural estimate of the characteristic force scale separating the weakly and strongly bound regimes, $f_{\rm{c}}\sim f^{*}$. In fact, the same reasoning also applies to the crossover estimation in the Gaussian slinky model. In the weakly bound limit ($w\ll1$), the free energy density [Eq.~\eqref{psi_FJN_w<1}] reduces to $\psi\approx-k_{\rm{B}}T\exp(-w^{-2})$, from which the crossover can be inferred to occur around $w\approx1$---the heuristic criterion used to separate the two asymptotic regimes---yielding the same estimate for the crossover force scale [Eq.~\eqref{f_c_necklace}].

The crossover estimation can also be formulated in terms of the localization length, defined as
\begin{equation}\label{l_localization}
    {l_{\perp}}^2=\frac{\int d^2r_{\perp}\,{r_{\perp}}^2{\lvert\phi_0(r_{\perp})\rvert}^2}{\int d^2r_{\perp}\,{\lvert\phi_0(r_{\perp})\rvert}^2},
\end{equation}
an important observable that characterizes the spatial extent of the bound state \cite{Nelson2}. Outside the potential well, the bound-state eigenfunction decays as $\phi_0(r_{\perp})\!\sim\!K_0(\kappa r_{\perp})$, exhibiting an exponential tail $\!\sim\!\exp(-\kappa r_{\perp})$ with the characteristic decay length $\kappa^{-1}$. In the weakly bound regime, the decay length greatly exceeds the well radius ($\kappa^{-1}\gg\sqrt{a/\pi}$), and the interior eigenfunction is $\phi_0(r_{\perp})\!\sim\!J_0(k r_{\perp})\approx1$. Consequently, the contribution from the interior to the integrals in Eq.~\eqref{l_localization} is parametrically negligible, and the dominant weight arises from the exterior region. Neglecting the interior piece and extending the lower limit to zero, it is then approximated as
\begin{equation*}
    {l_{\perp}}^2\approx\frac{\int_0^{\infty}d^2r_{\perp}\,{r_{\perp}}^2{\lvert K_0(\kappa r_{\perp})\rvert}^2}{\int_0^{\infty}d^2r_{\perp}\,{\lvert K_0(\kappa r_{\perp})\rvert}^2}=\frac{2}{3}\kappa^{-2}.
\end{equation*}
Thus, the localization length is approximately given by
\begin{equation}\label{l_localization~kappa}
    l_{\perp}\approx\kappa^{-1}\approx\sqrt{\frac{a}{\pi}}\,\exp\!\left[2\!\left(\frac{f^*}{f}\right)\right],
\end{equation}
which decreases exponentially as the force increases. From the spatial perspective, the crossover occurs when the localization length becomes comparable to the geometric scale of the potential well, $l_{\perp}\sim\sqrt{a/\pi}$, yielding $f_{\rm{c}}\sim f^{*}$.

Notably, the crossover force scale $f_{\rm{c}}$ for the shallow-well case captured by the quantum analogy exhibits the same functional form as that obtained from the Gaussian slinky model [Eq.~\eqref{f_c_necklace}], upon identifying
\begin{equation*}
    \varepsilon_a=k_{\rm{B}}T\ln\left[1+\exp\left(\frac{\varepsilon}{k_{\rm{B}}T}\right)\right]\approx k_{\rm{B}}T\exp\left(\frac{\varepsilon}{k_{\rm{B}}T}\right).
\end{equation*}
Although the two approaches yield the same scaling behavior of $f_{\rm{c}}$, they apply to different parameter ranges, leading to dramatically different magnitudes. For instance, using the same parameter values as in Fig.~\ref{fig:n_c_WLN} but assuming a shallow-well case with $\varepsilon_a/k_{\rm{B}}T\approx0.3$, we find $f_{\rm{c}}\sim\SI{100}{\nano\newton}$, which lies far beyond the force scale accessible within the Gaussian slinky model (piconewtons). The two approaches are therefore complementary in their respective regimes of validity, each compensating for the limitation of the other: the Gaussian slinky model is constrained by the accessible force range but not by the confining potential depth, whereas the quantum analogy within our continuum-limit description is constrained by the potential depth but not by the force.

The crossover force may likewise be estimated by matching the binding free energy densities of the weakly and strongly bound regimes, as in the previous section [Eq.~\eqref{crossover}]. This method also yields the same result, $f_{\rm{c}}\sim f^{*}$, and is presented in Appendix~\ref{app:B}.

We point out that the crossover force can be estimated using a simple mean-field argument, without recourse to the quantum analogy. To this end, we revisit the continuum-limit setup (Fig.~\ref{fig:CL}) using a more compact notation: consider a directed polymer confined by a columnar pinning potential with linear energy density $V_0$ and radius $\Delta$. Under strong confinement (localization), the thermal energy $k_{\rm{B}}T$ is comparable to the binding energy over a longitudinal segment of length $k_{\rm{B}}T/V_0$, and the slope of the tilting fluctuations is $\Delta/(k_{\rm{B}}T/V_0)$. The unbinding crossover occurs when the thermal energy exceeds the corresponding tilting energy:
\begin{equation}\label{crossover_mean-field}
    k_{\rm{B}}T>\frac{f\Delta^2}{2\,(k_{\rm{B}}T/V_0)}.
\end{equation}
This inequality can be written as $f<f_{\rm{c}}$, where
\begin{equation}\label{f_c_mean-field}
f_{\rm{c}}=\frac{2(k_{\rm{B}}T)^2}{V_0\Delta^2},   
\end{equation}
which is identical to the result obtained from the quantum analogy (up to an insignificant numerical prefactor of 2). However, this mean-field argument does not reveal whether delocalization occurs through a sharp phase transition or as a smooth crossover. In contrast, the quantum analogy demonstrates that a bound state persists even for an arbitrarily shallow potential well, implying the absence of any phase transition. Moreover, it provides a quantitative measure of the bound-state spatial extent---the localization length [Eq.~\eqref{l_localization}].

\subsection{Necklace--slinky correspondence}

In establishing the mapping between the bound cross-links and the emergent loops in the projected plane, we have implicitly assumed that the reversible cross-linking in the necklace is equivalent to the reversible looping in the Gaussian slinky representation. However, the underlying binding mechanisms differ between the two models. In the necklace model, the cross-linking interaction is described by a harmonic potential characterized by the spring constant $\xi$ and the chemical potential (or activation energy) $\mu$, whereas in the Gaussian slinky model, the looping interaction is represented by a confining potential well of effective range $a$ and depth $\varepsilon_a$ (or equivalently, $\varepsilon$). This naturally gives rise to the question of how the two implementations can be reconciled, that is, how the different parameterizations of a fundamentally single phenomenon are connected in this mapping. We address this question by comparing the statistical weights associated with binding in each model.

In the Gaussian slinky, the partition function restricted to looped configurations is identified with the looping probability, which serves as the statistical weight of the loop [see Eq.~\eqref{Z_GL}]. In the language of the necklace model, the looping probability corresponds to the probability of cross-link formation. Building on the results from Secs.~\ref{sec2} and~\ref{sec3}, we can evaluate the cross-linking probability and equate it with the looping probability. The resulting relation creates a bridge between the two descriptions. Let us first consider the cFJL of size $N_0$ in which the single cross-link is now reversible, as in the FJN. Using Eqs.~\eqref{Z_cFJL} and~\eqref{Z_FJC}, the partition function for the uncross-linked state follows as
\begin{equation}\label{Z_u}
    \mathscr{Z}_{\rm{u}}=\left(\mathscr{Z}_{\rm{FJC}}\right)^2-\mathscr{Z}_{\rm{cFJL}},
\end{equation}
and the partition function for the cross-linked state is given by
\begin{equation}\label{Z_c}
    \mathscr{Z}_{\rm{c}}=\mathscr{Z}_{\rm{cFJL}}\exp\left(\frac{\mu}{k_{\rm{B}}T}\right).
\end{equation}
The total partition function for the reversible cFJL reads $\mathscr{Z}_{\rm{u}}+\mathscr{Z}_{\rm{c}}$, and the cross-linking probability is obtained as the relative weight of the cross-linked state normalized by the total. For $N_0\gg1$, the probability is given by
\begin{equation}\label{P_c}
    P_{\rm{c}}=\frac{\mathscr{Z}_{\rm{c}}}{\mathscr{Z}_{\rm{u}}+\mathscr{Z}_{\rm{c}}}\approx\frac{f}{2N_0\xi b}\exp\left(\frac{\mu}{k_{\rm{B}}T}\right).
\end{equation}
In the projected representation of the reversible cFJL, the corresponding looping probability is $w$, as given in Eq.~\eqref{w}. Finally, equating $P_{\rm{c}}=w$ yields the following relation between the model parameters:
\begin{equation}\label{parameter_relation}
    \xi\exp\left(-\frac{\mu}{k_{\rm{B}}T}\right)=\frac{2\pi k_{\rm{B}}T}{a}\exp\left(-\frac{\varepsilon}{k_{\rm{B}}T}\right).
\end{equation}

The wormlike case also satisfies the same parameter relation, as expected on physical grounds, yet still noteworthy for its consistency. Analogous to Eqs.~\eqref{Z_u} and~\eqref{Z_c}, the partition functions of the reversible cWLL in each state are constructed from Eqs.~\eqref{Z_cWLL} and~\eqref{Z_WLC}:
\begin{equation}\label{Z_u'}
    {\mathscr{Z}_{\rm{u}}}^{\prime}=\left(\mathscr{Z}_{\rm{WLC}}\right)^2-\mathscr{Z}_{\rm{cWLL}},
\end{equation}
\begin{equation}\label{Z_c'}
    {\mathscr{Z}_{\rm{c}}}^{\prime}=\mathscr{Z}_{\rm{cWLL}}\exp\left(\frac{\mu}{k_{\rm{B}}T}\right).
\end{equation}
Using the effective bond length $\widetilde{b}=2l_{\rm{m}}$, the cross-linking probability for $N_0=L_0/\widetilde{b}\gg1$ then takes the same form as Eq.~\eqref{P_c}:
\begin{equation}\label{P_c'}
    {P_{\rm{c}}}^{\prime}=\frac{{\mathscr{Z}_{\rm{c}}}^{\prime}}{{\mathscr{Z}_{\rm{u}}}^{\prime}+{\mathscr{Z}_{\rm{c}}}^{\prime}}\approx\frac{f}{2N_0\xi\widetilde{b}}\exp\left(\frac{\mu}{k_{\rm{B}}T}\right).
\end{equation}
Similarly, equating this probability with the corresponding looping probability $w^{\prime}$, given in Eq.~\eqref{w'}, recovers the relation in Eq.~\eqref{parameter_relation}. The bridge between the harmonic and confining descriptions of binding thus remains robust across different chain flexibilities. Such robustness can be attributed to the fact that the cross-linking interaction itself is independent of the chain's flexibility, while the looping interaction is consistently embedded within the Gaussian framework.

This universal relation directly links the cross-linking parameters ($\xi,\mu$) in the necklace model to the looping parameters ($a,\varepsilon$) in the Gaussian slinky representation. First, the chemical potential (or activation energy) $\mu$ conceptually maps onto the energy parameter $\varepsilon$, as both set the energetic cost opposing the entropic drive toward unbinding. Naturally, the spring constant of the cross-link and the areal range of the looping interaction are related by $\xi=2\pi k_{\rm{B}}T/a$. In general, the spring constant for a two-state system with a reversible harmonic cross-link can be deduced by constructing an effective Gaussian potential~\cite{PB2}, though the approximation holds only in the weak-binding regime ($\mu<0$). Employing this method leads to the slightly different expression $\xi=2k_{\rm{B}}T/a$, based on identifying the effective interaction range of the Gaussian potential with twice its variance. Both approaches yield essentially the same result for the spring constant. The effective Gaussian potential method merely shifts the mapping from $(\mu\longleftrightarrow\varepsilon)$ to $(\mu\longleftrightarrow\varepsilon-k_{\rm{B}}T\ln\pi)$, without altering the qualitative correspondence between the parameters. The existence of these correspondences reinforces the internal coherence of the mapping between the necklace system and its projected Gaussian representation.

\section{Conclusions}\label{sec5}

In the first part of this article, we analyze the tensile elasticity of two polymer chains that share one common end and are cross-linked at the other, modeling the cross-link as a harmonic potential. Our approach is purely analytic and applies to the weakly bending approximation. For a pair of freely jointed chains, this approximation is valid under strong tensile force, whereas for a pair of wormlike chains, it holds under strong stretching or when the persistence length is large. In either case, a transverse potential suffices, as the longitudinal contribution is of higher order in the bending angle. 

In all cases of strong stretching that we examined, the cross-link contribution to the force--extension relation takes the same functional form. Remarkably, this contribution is always positive, implying that the cross-link induces additional stretching of the system. This behavior arises because the cross-link restricts the transverse fluctuations of the two chains. However, in most cases, the magnitude of the effect is negligible. Under strong stretching, the cross-link contribution decays as $\sim f^{-2}$, which is subdominant compared with the $\sim f^{-1}$ behavior of the FJC or the $\sim f^{-1/2}$ behavior of the WLC. In the thermodynamic limit, the contribution vanishes as $\sim(Nf)^{-1}$ [or $\sim(Lf)^{-1}$], as expected from a lever-arm argument. At the strong $\xi$ limit, a noticeable effect remains only for small $N$ in the freely jointed case; otherwise, the thermodynamic limit takes over and the contribution becomes negligible. We note that the small $N$ regime is not merely a formal limit, but could be realized experimentally (at least in principle) using DNA nunchuck nanotechnology \cite{Fygenson_Nano}, and related studies on hinged, rodlike polymers has been reported in Refs.~\cite{PB_Hinge_PRE,PB_Razbin_Nunchucks}. Another noteworthy result in the strong stretching regime is that the differential stretching compliance of the loop (or simply the pair, as the cross-link becomes irrelevant) is, for any given force, half that of a single chain under the same conditions, reminiscent of two springs connected in parallel. 

In the case of a pair of WLCs in the rodlike limit under moderate stretching, the cross-link can have a significant effect on the force--extension relation when $\xi$ is large. However, in this regime the two chains essentially fluctuate sliding past each other. In order to get a \textit{bona fide} loop for rodlike polymers, we need to go beyond the Gaussian (weakly bending) approximation and include the longitudinal component of the harmonic cross-linking potential.

Even though the cross-link has a negligible effect on the force--extension relation in the regimes we considered, its effect on reducing the transverse mismatch between the two chains is substantial and becomes even more pronounced in the thermodynamic limit. This reduction justifies the use of a transverse harmonic spring that effectively forms a loop out of two parallel, stretched chains. We have explicitly calculated both the transverse and longitudinal mismatches of the chain end-points for all systems that we examined.

In the second part of the article, we consider a pair of strongly stretched polymers (FJCs or WLCs), in the thermodynamic limit (polymer necklace), with a regular sequence of two-state cross-linking sites along their contour. Each cross-linking site can be either occupied or unoccupied. We show that a reversible cross-link, modeled as a harmonic spring whose on-off probability is controlled by a chemical potential, can be mapped to a potential well of finite depth and range. Using the generating function method, we calculate the mean fraction of occupied sites as a function of the stretching force, for both the freely jointed and the wormlike necklace. We obtain analytic expressions in two asymptotic regimes: when the force is strong but not too strong so that the mean occupation number remains small, and when the force is so strong that the mean occupation number is close to one. The resulting behavior implies a sigmoidal dependence of the mean occupation number on the applied tension. There are two regimes: the curve is concave at weaker tension and convex at stronger tension. The theory of the necklace model clearly states that these two regimes are connected by a crossover rather than a phase transition, the latter being excluded by the scaling exponent associated with the large uncross-linked blocks (bubbles) in the system.

Our calculation is based on what we call the Gaussian slinky. Noting that the transverse projection of a strongly stretched FJC (with a large degree of polymerization) or WLC behaves as a Gaussian chain, we map the polymer necklace to a two-dimensional system of concatenated, reversible  Gaussian loops of various sizes, where two Gaussian strands of equal degree of polymerization get (reversibly) linked at their ends to form a loop. This description breaks down when the stretching force becomes so large that the Gaussian loop size in the slinky is comparable to the range of the cross-link. This force regime may not be very interesting, if almost all the cross-links are already bound, or it may be interesting, if most cross-links remain unbound even at the strong force limit of the model. The latter situation could happen when the binding potential depth is small compared with $k_{\rm{B}}T$.

In order to address the question about the strong force behavior of our system for shallow binding potentials, we consider a continuum limit in which the discrete cross-linking sites are replaced by a coarse-grained, linear potential well. This continuum model allows us to invoke the analogy between the classical statistical mechanics of a directed polymer in a columnar pin in $(1+2)$ dimensions and the quantum mechanics of a particle in two dimensions subject to a 2D potential well. The quantum analogy reveals two binding regimes, separated by a crossover. In the strongly bound regime, the two chains fluctuate, on average, within the range of the potential; in the weakly bound regime, they fluctuate at an average distance much larger than the potential range. This average transverse distance corresponds to the localization length that can easily be inferred from the quantum analogy. We have determined the crossover force in three different ways: from the localization length, from the localization free energy, and from a mean-field argument that does not rely on the quantum analogy. However, we need the quantum analogy to ensure that the crossover is not a genuine phase transition and the two chains always remain bound to each other.

Our work can be extended in several interesting directions. For example, we may consider bundles formed by more than two reversibly cross-linked, strongly stretched chains. The case of infinitely many parallel-aligned, reversibly cross-linked directed polymers was treated in Ref.~\cite{Dutta_bundle_EPL_2016}. Another interesting direction for future work is the study of strongly stretched loops and bundles subject to a pulling force acting along the entire contour of the strands, as in the case of flow. The statics of such pulled loops was examined in Ref.~\cite{Julicher_pulled_loop_PRL_2015} and their dynamics was studied in Ref.~\cite{Julicher_pulled_loop_NJP_2018}.

\begin{acknowledgments}
    The early stage of this work was supported by the National Research Foundation of Korea (NRF) under Grant No.~NRF-2022R1F1A1070341, funded by the Ministry of Science and ICT (MSIT), Korea.
\end{acknowledgments}

\appendix

\section{SINGLE WORMLIKE CHAIN}\label{app:A}

For completeness, we present here the results for a single wormlike chain (WLC) within the same framework as used in the cross-linked wormlike loop (cWLL). For the single system, the effective Hamiltonian reads
\begin{equation}
    \mathscr{H}_{\rm{WLC}}=\int_{0}^{L}ds\left[\frac{\kappa}{2}\left(\frac{\partial\bm{t}(s)}{\partial s}\right)^2-ft_{x}(s)\right].
\end{equation}
In the weakly bending regime, where the approximation in Eq.~\eqref{t_i,x} holds, the Hamiltonian becomes
\begin{equation}
    \mathscr{H}_{\rm{WLC}}=\frac{1}{2}\int_{0}^{L}ds\left[\kappa\left(\frac{\partial\bm{t}_{\perp}(s)}{\partial s}\right)^2+f{\bm{t}_{\perp}(s)}^2\right]-fL.
\end{equation}
Since the transverse directions decouple symmetrically, the analysis reduces to a single-component partial Hamiltonian
\begin{equation}\label{H_WLC*}
    \mathscr{H}_{\rm{WLC}}^{*}=\frac{1}{2}\int_{0}^{L}ds\left[\kappa\left(\frac{\partial{t}_{\perp}(s)}{\partial s}\right)^2+f{t_{\perp}(s)}^2\right].
\end{equation}
Consistent with the boundary conditions of vanishing bending moment [Eq.~\eqref{bc}], $t_{\perp}(s)$ can be expanded in a Fourier cosine series
\begin{equation}\label{t_cosine}
        t_{\perp}(s)=\frac{c_{0}}{2}+\sum_{m=1}^{M}c_{m}\cos(k_{m}s),
\end{equation}
where $k_{m}\equiv m\pi/L$. Substituting this expansion into Eq.~\eqref{H_WLC*} gives the Hamiltonian in Fourier space
\begin{equation}\label{H_WLC*_Fourier}
    \mathscr{H}_{\rm{WLC}}^{*}=\frac{Lf{c_{0}}^2}{8}+\frac{L}{4}\sum_{m=1}^{M}\left(\kappa{k_{m}}^2+f\right){c_{m}}^2,
\end{equation}
and the corresponding partition function is well approximated by a product of Gaussian integrals, yielding
\begin{equation}
    \mathscr{Z}_{\rm{WLC}}^{*}=\sqrt{\frac{8\pi k_{\rm{B}}T}{Lf}}\prod_{m=1}^{M}\sqrt{\frac{4k_{\rm{B}}TL/\pi\kappa}{m^2+\alpha^2}},
\end{equation}
where $\alpha\equiv\sqrt{L^2f/\pi^2\kappa}$. The complete partition function is then assembled as
\begin{equation}\label{app:Z_WLC}
    \mathscr{Z}_{\rm{WLC}}=\frac{8\pi\exp(Lf/k_{\rm{B}}T)}{Lf/ k_{\rm{B}}T}\prod_{m=1}^{M}\frac{4k_{\rm{B}}TL/\pi\kappa}{m^2+\alpha^2}.
\end{equation}

The average extension of the WLC is given by
\begin{equation}\label{X-f_WLC}
    \langle X\rangle_{\rm{WLC}}=L-\frac{k_{\rm{B}}T}{f}-\sum_{m=1}^{M}\frac{k_{\rm{B}}TL^2/\pi^2\kappa}{m^2+\alpha^2}.
\end{equation}
Assuming $M\gg1$, the infinite-series approximation combined with the series expansion of the coth function yields the closed-form expression
\begin{equation}\label{app:X-f_WLC_closed}
    \langle X\rangle_{\rm{WLC}}=L-\frac{k_{\rm{B}}TL}{\sqrt{4\kappa f}}\coth\left(\sqrt{\frac{L^2f}{\kappa}}\right)-\frac{k_{\rm{B}}T}{2f}.
\end{equation}
For $l_{\rm{m}}=\sqrt{\kappa/f}\ll L$, the average extension is approximated by
\begin{equation}\label{X-f_WLC_ordinary}
    \langle X\rangle_{\rm{WLC}}\approx L-\frac{k_{\rm{B}}TL}{\sqrt{4\kappa f}}-\frac{k_{\rm{B}}T}{2f},
\end{equation}
and for $l_{\rm{m}}\gg L$, it becomes
\begin{equation}\label{X-f_WLC_rodlike}
    \langle X\rangle_{\rm{WLC}}\approx L-\frac{k_{\rm{B}}TL^2}{6\kappa}-\frac{k_{\rm{B}}T}{f}.
\end{equation}

It is noteworthy that if clamped boundary conditions,
\begin{equation}
    t_{\perp}(s)\vert_{s=0}=t_{\perp}(s)\vert_{s=L}=0,
\end{equation}
are applied instead of the free-end conditions [Eq.~\eqref{bc}], they lead to a Fourier sine representation
\begin{equation}\label{t_sine}
    t_{\perp}(s)=\sum_{m=1}^{M}c_m\sin(k_m s),
\end{equation}
in place of the cosine series [Eq.~\eqref{t_cosine}]. As a result of this change, the average extension is also modified as
\begin{equation}\label{X-f_WLC_clamped}
    \langle X\rangle_{\rm{WLC}}\Big|_{\rm{clamped}}=\langle X\rangle_{\rm{WLC}}\Big|_{\rm{free}}+\frac{k_{\rm{B}}T}{f},
\end{equation}
where $\langle X\rangle_{\rm{WLC}}|_{\rm{free}}$ denotes the free-end result. This marginally larger extension, accompanied by a stiffer response, reflects the fact that the clamping suppresses bending fluctuations near the ends, reducing overall bending energy. However, such boundary conditions are incompatible with the cWLL model, where the endpoint, and thus the cross-link, is allowed to fluctuate.

The longitudinal fluctuations of the WLC are given by
\begin{align}\label{fluctuation_l_WLC}
    &\left\langle\left(R_{\parallel}-\langle R_{\parallel}\rangle\right)^2\right\rangle_{\rm{WLC}}=\frac{{k_{\rm{B}}}^2T^2L^2}{4\kappa f}\csch^2\left(\sqrt{\frac{L^2f}{\kappa}}\right)\nonumber\\
    &\quad+\frac{{k_{\rm{B}}}^2T^2L}{\sqrt{16\kappa f^3}}\coth\left(\sqrt{\frac{L^2f}{\kappa}}\right)+\frac{{k_{\rm{B}}}^2T^2}{2f^2},
\end{align}
in the limit $M\gg1$. Unlike the longitudinal fluctuations, the transverse fluctuations cannot be obtained directly from derivatives of $\ln\mathscr{Z}_{\rm{WLC}}$. This marks a subtle difference, as such a derivation is possible in the more complex cWLL model. Yet they can still be obtained via the equipartition theorem. 

The transverse fluctuations can be expressed as the ensemble average
\begin{equation}
    \langle{\bm{R}_{\perp}}^2\rangle_{\rm{WLC}}=\left\langle\left(\int_{0}^{L}ds\ \bm{t}_{\perp}(s)\right)^2\right\rangle.
\end{equation}
Substituting Eq.~\eqref{t_cosine} into this expression yields
\begin{equation}\label{fluctuation_t_WLC_Fourier}
    \langle{\bm{R}_{\perp}}^2\rangle_{\rm{WLC}}=2\times\frac{L^2}{4}\left\langle{c_0}^2\right\rangle=\frac{L^2}{2}\left\langle{c_0}^2\right\rangle.
\end{equation}
Since the Hamiltonian in Eq.~\eqref{H_WLC*_Fourier} is quadratic in the Fourier coefficients, the average energy of each mode can be determined from the equipartition theorem. For the zero mode ($n=0$),
\begin{equation}\label{E_0}
    \langle E_0\rangle=\frac{Lf}{8}\left\langle{c_0}^2\right\rangle=\frac{1}{2}k_{\rm{B}}T,
\end{equation}
and for the $n$-th mode ($n=1,2,3,\cdots$),
\begin{equation}\label{E_n}
    \langle E_n\rangle=\frac{L(\kappa{k_n}^2+f)}{4}\left\langle{c_n}^2\right\rangle=\frac{1}{2}k_{\rm{B}}T.
\end{equation}
Combining Eqs.~\eqref{fluctuation_t_WLC_Fourier} and~\eqref{E_0}, the transverse fluctuations are given by
\begin{equation}\label{fluctuation_t_WLC}
    \langle{\bm{R}_{\perp}}^2\rangle_{\rm{WLC}}=\frac{2k_{\rm{B}}TL}{f}.
\end{equation}

The average amplitude of the transverse fluctuations along the chain's contour can be written as
\begin{equation}\label{app:fluctuation_t_WLC_avg_1}
    \frac{1}{L}\int_{0}^{L}ds\left\langle{\left[{\bm{R}_{\perp}}(s)\right]}^2\right\rangle=2\times\frac{1}{L}\int_{0}^{L}ds\left\langle{\left[{R_{\perp}}(s)\right]}^2\right\rangle,
\end{equation}
where ${R_{\perp}}(s)\equiv\int_0^sds^{\prime}t_{\perp}(s^{\prime})$ is one Cartesian component of the local transverse fluctuations ${\bm{R}_{\perp}}(s)$ in the plane perpendicular to the force axis. Substituting Eq.~\eqref{t_cosine} for $t_{\perp}(s^{\prime})$ and applying the equipartition results given in Eqs.~\eqref{E_0} and~\eqref{E_n} yields
\begin{equation}\label{app:fluctuation_t_WLC_avg_2}
    \frac{1}{L}\int_{0}^{L}ds\left\langle{\left[{\bm{R}_{\perp}}(s)\right]}^2\right\rangle=\frac{2k_{\rm{B}}TL}{3f}+\sum_{m=1}^{M}\frac{2k_{\rm{B}}TL^3/\pi^4\kappa}{m^2(m^2+\alpha^2)}.
\end{equation}
Evaluating the sum with the identity
\begin{equation*}
    \sum_{m=1}^{\infty}\frac{1}{m^2(m^2+\alpha^2)}=\frac{\pi^2\alpha^2-3\pi\alpha\coth(\pi\alpha)+3}{6\alpha^4}
\end{equation*}
gives the closed-form expression for the contour-averaged transverse amplitude
\begin{align}\label{app:fluctuation_t_WLC_avg_3}
    \frac{1}{L}\int_{0}^{L}ds\left\langle{\left[{\bm{R}_{\perp}}(s)\right]}^2\right\rangle&=\frac{k_{\rm{B}}TL}{f}-\frac{k_{\rm{B}}Tl_{\rm{m}}}{f}\coth\left(\frac{L}{l_{\rm{m}}}\right)\nonumber\\
    &\quad+\frac{k_{\rm{B}}T{l_{\rm{m}}}^2}{Lf}.
\end{align}

\section{CROSSOVER ESTIMATION VIA MATCHING IN THE QUANTUM ANALOGY}\label{app:B}

For methodological consistency with the Gaussian slinky analysis in Sec.~\ref{sec:crossover} [Eq.~\eqref{crossover}], we estimate here the crossover force within the quantum-analogy framework by matching the binding free energy densities of the weakly and strongly bound regimes. In the main text, two approaches based on the asymptotic form of the binding free energy density and on the localization length in the weakly bound regime are used to estimate the crossover force. The present method follows the same spirit as those analyses.

In the strongly bound regime---which can be realized under strong stretching even for a shallow potential well---the binding free energy density approaches the well depth, $\lvert \lambda_0\rvert\approx\varepsilon_a/L_0$. In this limit, the finite depth of the potential becomes irrelevant, and the bound state behaves as if confined by an effectively hard-wall potential. Consequently, the corresponding quantum problem reduces to that of finding the ground-state energy of an infinitely deep circular potential well in two dimensions \cite{Robinett1,Robinett2}. The binding free energy density in the strongly bound regime is then obtained as
\begin{align}\label{g_s}
    g_{\rm{s}}&\approx-\frac{\varepsilon_a}{L_0}+\frac{\pi{j_{0,1}}^2{k_{\rm{B}}}^2T^2}{2\widetilde{f}a}\nonumber\\
    &\approx-\frac{\varepsilon_a}{L_0}\exp\left(-\frac{\pi{j_{0,1}}^2{k_{\rm{B}}}^2T^2L_0}{\varepsilon_aaf}\right),
\end{align}
where $j_{0,1}$ is the first zero of the Bessel function $J_0(x)$. The second approximation expresses the small correction term in exponential form, easing the comparison with the weakly bound result.

The crossover between the weakly and strongly bound regimes is characterized by a range of forces around $f_{\rm{c}}$, where the two binding free energy densities become comparable, $g_{\rm{w}}(f_{\rm{c}})\approx g_{\rm{s}}(f_{\rm{c}})$. This condition can be expressed compactly as
\begin{equation}\label{Lambert}
    \eta\chi_{\rm{c}}\exp(\eta\chi_{\rm{c}})=\eta,
\end{equation}
where $\eta\equiv{j_{0,1}}^2-4$ and $\chi_{\rm{c}}\equiv\pi k_{\rm{B}}^2T^2L_0/\varepsilon_aaf_{\rm{c}}$. The solution of this transcendental equation is given in terms of the Lambert~W function as $\chi_{\rm{c}}=W(\eta)/\eta$ \cite{Lambert}. With $j_{0,1}\approx2.4$, we obtain a practical estimate for the crossover force
\begin{equation}\label{f_c_quantum}
    f_{\rm{c}}\approx\frac{1.8\pi{k_{\rm{B}}}^2T^2L_0}{W(1.8)\varepsilon_aa}\approx\frac{2.2\pi{k_{\rm{B}}}^2T^2L_0}{\varepsilon_aa}\sim\frac{\pi{k_{\rm{B}}}^2T^2L_0}{\varepsilon_aa}.
\end{equation}
Thus, the matching condition reproduces the same crossover estimation as that obtained in Sec.~\ref{sec:quantum} [Eq.~\eqref{f_characteristic}], confirming the consistency among the different approaches.

\bibliography{ref.bib}

\end{document}